# Can Institutional Integration of Western Balkans' Stock Exchanges Strengthen Monetary Transmission?

Stefan Tanevski
stefan.tanevski@uacs.edu.mk
University American College Skopje

# Abstract

This paper asks how institutional stock-market integration reshapes the transmission of monetary policy through asset prices in small open economies. Motivated by the persistent segmentation of Western Balkan capital markets, we develop a two-stage counterfactual transmission framework to identify how stock-exchange consolidation would alter the elasticity of market valuations to monetary shocks.

First, a synthetic-control simulation constructs a counterfactual integrated Western Balkan stock exchange comprising Bosnia and Herzegovina, North Macedonia, and Serbia, benchmarked to the Baltic OMX merger, thereby quantifying the structural valuation gains of institutional integration. Second, we identify exogenous monetary-policy innovations using a Taylor-rule framework augmented with inflation and output forecasts and reserve adjustments. These shocks are then embedded within a Local-Projections estimator à la Jordà (2005) to trace the dynamic responses of market capitalisation under fragmented and integrated market regimes.

The results point to a systematic amplification of monetary-policy transmission through the asset-price channel once markets are unified. Following a policy tightening of about 100 basis points, equity valuations fall roughly twice as strongly under integration than under fragmented markets. Additionally, we find that integration alters the sensitivity of monetary transmission itself: the initial pass-through intensifies, but its marginal responsiveness to further integration declines over time, signalling the consolidation of a new steady-state regime.

Table of Contents

# List of Figures

# List of Tables

1. Introduction

The fragmentation of equity markets in the Western Balkans remains one of the most persistent structural weaknesses of the region's financial architecture. Despite two decades of institutional transition and broader integration efforts such as the Stabilisation and Association Process, CEFTA, the Common Regional Market, and the Open Balkans Initiative, financial integration has lagged behind other dimensions of economic convergence. While trade and banking linkages have become increasingly cross-border, domestic equity markets remain nationally segmented, shallow, and institutionally isolated. Regulatory frameworks have largely converged toward the European acquis, yet the region still lacks a unified framework for investment, intermediation, and risk-sharing. As a result, domestic exchanges are too small to support meaningful diversification, cross-border trading remains minimal, and capital markets play only a limited role in transmitting macro-financial signals. This structural gap limits both financial stability and monetary-policy traction, renderring these economies bank-centric.

Experience from the early 2000s in Europe suggests that stock-exchange consolidation can generate gains that extend beyond micro-level efficiency improvements. The merger of the Baltic exchanges into the OMX system in 2004, along with similar initiatives in Central and Eastern Europe, was followed by a sustained deepening of liquidity, lower transaction costs, and enhanced information efficiency (Jazepcikaite, 2008; Dorodnykh, 2013). These outcomes motivate a broader question: can institutional stock-market integration also reshape the macroeconomic transmission mechanism, i.e., the elasticity of asset prices to monetary shocks? In principle, integrated markets should transmit policy signals more swiftly and price information more efficiently, thereby strengthening the asset-price channel of monetary policy (Ehrmann and Fratzscher, 2006; Meier, 2013; Gali, 2015). In this sense, institutional integration of stock exchanges becomes not only a financial-development strategy but also a macroeconomic transmission reform, linking market structure to policy effectiveness.

In the Western Balkan context, however, this hypothesis remains untested. No regional exchange merger has taken place, and the data do not contain a clear policy event that would allow for standard identification strategies such as difference-in-differences or event studies. Moreover, heterogeneous institutional quality and partial euroisation complicate causal inference. Addressing these challenges requires a framework capable of constructing a credible counterfactual for exchange integration and tracing its implications for monetary-policy transmission.

This paper develops such a framework through what we term a two-stage counterfactual transmission design. In the first stage, we employ the Synthetic Control Method (SCM) (Abadie and Gardeazabal, 2003; Abadie et al., 2010) to simulate how the market capitalisation of a hypothetical Western Balkans exchange comprising Serbia, Bosnia and Herzegovina, and North Macedonia (WB3) would have evolved had these markets consolidated. The Baltic OMX merger serves as an institutional and quantitative benchmark for scale and sequencing.

In the second stage, we estimate Taylor-rule-based monetary-policy shocks, incorporating forward-looking inflation and output forecasts, as well as official-reserve adjustments, to reflect the hybrid and partially euroised policy regimes typical of the region. These shocks are then embedded in a Local Projections (LP) model à la Jordà (2005) to assess how monetary-policy innovations affect equity valuations under two counterfactual configurations, i.e., fragmented national exchanges and a unified Western Balkans (WB3) stock exchange.

This combined SCM–LP framework enables a dual analysis, i.e., it captures both the static valuation effects of integration, changes in market capitalisation levels, and the dynamic transmission effects, i.e., how monetary shocks propagate over time under integration. To capture these dynamics, we introduce an integration-sensitivity coefficient ($g_h$), a horizon-dependent semi-elasticity that measures how the responsiveness of market capitalisation evolves as financial linkages deepen. This parameter provides a new empirical measure of transmission elasticity, typically absent in existing SCM or LP applications when conducted separately.

The results point to a consistent narrative. The synthetic simulations indicate that a hypothetical WB3 merger of stock exchanges would have produced a moderate yet persistent uplift in market capitalisation, interpreted as a credibility-driven and structurally anchored gain rather than a short-lived valuation surge. Building on this, the Local Projections results show that integration strengthens the asset-price channel of monetary policy, i.e., a 100-basis-point tightening produces a deeper and more persistent decline in equity valuations under integration, with responsiveness increasing by roughly 30–50 percent at peak horizons. Integration therefore alters both the magnitude and the slope of adjustment.

To our knowledge, this is the first paper to quantify the effects of a prospective stock-exchange merger in the Western Balkans and to empirically test the transmission of monetary-policy shocks within a counterfactual integrated-market environment. By combining SCM and LP within a unified empirical framework, the paper bridges institutional-finance and monetary-transmission literatures, offering an applied link between financial integration and macroeconomic performance in transition economies.

The implications are twofold. First, institutional consolidation of exchanges can yield durable gains in credibility, transparency, and investor confidence, provided that sufficient absorptive capacity and regulatory coherence exist. Second, financial integration can serve as a monetary-transmission reform, enhancing the responsiveness of financial variables to policy stance and improving the efficiency of information diffusion. In this sense, structural integration reinforces both market depth and policy traction, strengthening the role of capital markets within the broader macro-financial framework of the Western Balkans.

The remainder of the paper is structured as follows. Section 2 reviews the literature and situates the analysis within the experience of transition economies. Section 3 outlines the methodological rationale for combining SCM and LP and describes the data. Section 4 presents the synthetic baselines and comparative benchmarks and analyses the simulated merger effects. Section 5 examines the monetary-policy transmission mechanism. Section 6 reports robustness diagnostics, and Section 7 concludes.

2. Literature Review

Stock exchange mergers have increasingly been explored within the broader discourse of financial market integration, particularly in the context of post-liberalization reforms and the shift from state-controlled to privatized financial systems. This transformation has drawn scholarly attention across several domains including the evolution of emerging stock markets (Claessens et al., 2000); legal, institutional, and regulatory conditions conducive to mergers (Dorodnykh, 2014); and comparative analyses of consolidation outcomes with respect to liquidity, efficiency, and market reach (Pagano et al., 2002; Pagano et al., 2005; Nielsen, 2009).

Initial debates around stock market consolidation emerged in the 1990s, primarily as a response to governance bottlenecks in mutual exchanges, that is, decision-making inefficiencies and conflicts of interest arising from member-owned governance structures. These exchanges, governed collectively by their members, were expected to balance the interests of diverse stakeholders, including brokers, dealers, and institutional investors. However, as financial markets became more complex and fragmented, the alignment of interests deteriorated, creating strategic deadlocks and operational inefficiencies (Steil & Aggarwal, 2002; Phillips et al., 2014; Aggarwal and Dahiya, 2006). In this context, the ownership and management of stock exchanges were often concentrated within the same membership structure, which blurred lines between governance, regulation, and profit-making. This structure became increasingly incompatible with the demands of modern, competitive capital markets.

The solution emerged through demutualisation, a structural transformation that separates ownership rights from trading privileges, converting exchanges into for-profit entities accountable to shareholders rather than exchange members (de Passos Homem de Sá, 2009). This shift entailed the creation of independent corporate boards tasked with maximizing firm value, and represented a turning point in the governance of exchanges. Demutualisation began to take hold globally in the late 1990s and early 2000s. The process implied the transformation of exchanges from member-owned cooperatives into shareholder-owned corporations, aiming to resolve conflicts of interest and enable more agile decision-making. Though initially met with resistance due to concerns over losing self-regulatory authority, demutualisation gained traction as revenues declined and the limitations of mutual governance became more apparent, particularly the inability to modernize trading infrastructure, attract external capital, and respond competitively to market liberalization pressures. These limitations were not externally imposed, but rather stemmed from the internal structure of mutual exchanges, particularly the inability to modernize trading infrastructure, attract external capital, and respond competitively to market liberalization pressures where collective ownership and conflicting interests among members often hindered strategic decision-making and innovation (Macey & O'Hara, 2002; Cospormac, 2009).

Demutualisation was further accelerated by the digitalisation of trading infrastructure, which rendered geographic borders obsolete and introduced product and process innovations (Domowitz & Steil, 1999). At the same time, regulatory liberalization, such as the EU's Investment Services Directive (ISD), enabled firms to cross-list, thereby reducing the primacy of national exchanges and fueling competition across jurisdictions (Ramos, 2006). This twin pressure technological and regulatory, intensified calls for consolidation, particularly among smaller exchanges seeking scale and sustainability (de Passos Homem de Sá, 2009).

In Europe, rising competition paradoxically contributed to a wave of mergers and integrations, as national exchanges sought to scale up and defend market relevance. These integrations took two forms: horizontal integration, through the consolidation of trading venues i.e., Euronext; and vertical integration, through the acquisition of brokerage firms and clearinghouses (Di Noia, 1998; Floreani and Polato, 2010). Although these strategies raise legitimate concerns about market concentration and monopolisation, proponents argue that the benefits in terms of reduced trading costs, increased liquidity, and economies of scale outweigh the risks, especially (Floreani and Polato, 2010; Dorodnykh, 2013). Simultaneously, the convergence of European economies under the Economic and Monetary Union (EMU) i.e., the breakdown of geographical borders, has been cited as a key driver of stock market integration, as the elimination of exchange rate risk and macroeconomic convergence mechanisms strengthened financial interlinkages across borders (Hardouvelis, 2007). Empirical evidence further supports this argument, with studies identifying unidirectional causality from the EMU and stock market integration, where macroeconomic harmonisation and financial development emerge as significant contributing factors (Kim et al., 2005).

Globalisation has reinforced these processes, expanding cross-border trading, clearing, and settlement activities (Dorodnykh, 2013; Dorodnykh, 2014). While these trends have arguably provided incentives for regional and international stock exchange consolidation, the degree to which integration has led to efficiency gains remains open for discussion. Beck et al. (2000) and Passad et al. (2003) suggest that deeper financial integration can enhance resource allocation and reduce macroeconomic volatility in developing countries, while Carrieri et al. (2007) identify liberalisation and regulatory reform as preconditions for successful integration.

The development of stock exchange mergers has unveiled in two phases. Initially, regional consolidations were more prevalent, exemplified by the formation of Euronext, the Nordic Exchange, and Bolsas y Mercados Españoles (BME), which integrated multiple national markets into single trading platforms (Dorodnykh, 2013; Polato & Floreani, 2010). In more recent years, integration has expanded beyond regional boundaries, with transatlantic mergers linking major European and American exchanges, such as the NYSE Group and Euronext. Beyond structural consolidation, stock exchange integration is widely associated with several financial and economic benefits, although the extent to which these materialise remains subject to empirical validation. Theoretical literature posits that increased market integration fosters higher liquidity, thereby reducing investment frictions and perceived risk. Enhanced liquidity facilitates smoother market entry and exit, allowing investors to acquire equity and divest their holdings swiftly and at minimal cost when immediate access to capital is needed. Additionally, liquid and deep equity markets are argued to improve resource allocation, facilitating long-term investment and potentially fostering higher economic growth. Furthermore, cross-border stock market integration is expected to lead to higher trading volumes, lower market volatility, and improved risk-sharing mechanisms (Dorodnykh, 2013). In addition, mergers and acquisitions among exchanges allow for the transfer of expertise, governance frameworks, and technological advancements between partner exchanges, thereby fostering institutional and operational efficiencies (Pagano et al., 2002; Gomes-Casseres et al., 2006; Hasan et al., 2010).

These structural transformations have also contributed to operational efficiencies, as mergers eliminate duplicative infrastructure, streamline access to liquidity pools, and reduce cost inefficiencies associated with fragmented trading environments (Floreani & Polato, 2010; Pagano et al., 2001). The resulting economies of scale reinforce a broader shift towards higher market efficiency and interconnected financial ecosystems (Dorodnykh, 2014). Standardisation of trading platforms plays a crucial role in reducing operational complexity, while increased liquidity and market depth make integration financially attractive for investors (McAndrews, 2002; Neuman et al., 2002). Moreover, financial intermediaries regard stock exchange integration as a mechanism for exploiting economies of scale, enhancing settlement efficiency, and broadening investment opportunities (Polato & Floreani, 2010). Despite the expansion of research in this area, most of the studies on stock exchange mergers remain theoretical, in fact an empirical comprehensive synthesis of the literature on stock exchange mergers remains absent, given that studies tend to focus on specific aspects in isolation rather than providing an integrated perspective; and impacts of mergers remain largely unknown even today. Presently, the literature on the effects of mergers appears to be scattered across distinct themes where only a handful of studies evaluate the outcomes of the mergers (Pagano and Padilla, 2005; Nielsen, 2009; Charles et al., 2014; Phillips et al., 2014).

For example, Pagano and Padilla (2005) investigate the economic and financial benefits of stock exchange integration using the case of Euronext, which consolidated the French, Belgian, Dutch, and Portuguese stock exchanges. Their findings suggest that stock exchange mergers yield substantial efficiency gains through cost savings, enhanced liquidity, and reduced trading costs. Significant cost reductions largely stem from infrastructural cost reductions, labor-shedding and a creation of single trading platform systems. Moreover, direct benefits to users included lower trading fees, increased market accessibility, and improved trading efficiency. The reduction in explicit trading costs (such as fees and commissions) and implicit costs (including bid-ask spreads) was not uniformly distributed across all participating exchanges. Pagano and Padilla (2005) highlight that Portugal experienced the lowest gains from integration. Slimane (2012) attributes this outcome to a size effect, suggesting that smaller exchanges may struggle to fully absorb the benefits generated by the merger. Moreover, Nielson (2009) examines the heterogeneous effects of the Euronext stock exchange merger on firm's liquidity using a fixed-effects regression with interaction terms for firm size and foreign sales. He finds that liquidity gains are concentrated among large firms and those with foreign sales, while small and domestic firms see no significant benefits. Nielsson (2009) explains that stocks with low trading volumes generally exhibit wider bid-ask spreads, resulting in lower liquidity. A stock exchange merger may expand the investor base, leading to a more active order book and a subsequent reduction in transaction costs. Furthermore, Nielsson (2009) finds that the Euronext merger strengthened its market share relative to the London Stock Exchange; however, there is no evidence to suggest that the merger improved Euronext's ability to attract new firm listings.

While these studies offer valuable insights into the micro-level effects of stock exchange mergers particularly in terms of trading costs, liquidity, and firm-level outcomes they do not address broader macro-financial implications. In particular, the literature remains silent on how such mergers may influence the transmission of monetary policy via the asset price channel. To our knowledge, no empirical study has compared the effectiveness of monetary policy transmission before and after a stock exchange consolidation, such as in the case of Euronext or OMX Nordics and Baltics.

**Table 1**. Key Empirical Studies on Stock Exchange Mergers

| Study | Focus | Method | Findings | Limitations |
|---|---|---|---|---|
| Claessens et al. (2000) | Emerging stock markets post-liberalization | Comparative institutional analysis | Stock market development shaped by transition from state to private ownership; institutional conditions critical. | Focuses on emerging markets generally, not mergers specifically. |
| Dorodnykh (2013, 2014) | Legal/regulatory conditions; integration outcomes | Theoretical and descriptive | Consolidation enhances efficiency; institutional alignment and governance are essential. | Lacks empirical testing in underdeveloped markets. |
| Pagano et al. (2002, 2005) | Euronext merger; efficiency and cost savings | Empirical evaluation | Substantial cost savings and liquidity improvements; uneven | Less applicable to smaller exchanges or |

| Study | Focus | Method | Findings | Limitations |
|---|---|---|---|---|
| | | | benefits across countries. | low-volume firms. |
| Nielsson (2009) | Firm-level liquidity effects in Euronext | Fixed-effects panel regression | Large and international firms gained liquidity; smaller/domestic firms saw no significant change. | Limited generalizability; case-specific to Euronext. |
| Steil & Aggarwal (2002); Aggarwal & Dahiya (2006); Phillips et al. (2014) | Governance inefficiencies in mutual exchanges | Descriptive case analysis | Mutual governance structures failed to meet modern market demands; demutualisation supported. | Primarily theoretical; lacks quantitative validation. |
| Hardouvelis (2007); Kim et al. (2005) | EMU effects on market integration | Macroeconomic causality tests | EMU promotes financial integration; macroeconomic convergence reinforces interlinkages. | Focused on macroeconomic linkages rather than micro-level exchange outcomes. |
| Beck et al. (2000); Carrieri et al. (2007) | Financial integration in developing countries | Theoretical + empirical support | Integration improves resource allocation and reduces volatility. | Does not focus specifically on stock exchange mergers. |
| Pownall et al. (2013) | Euronext institutional harmonization | Descriptive institutional study | NSC trading platform and harmonized rulebook facilitated integration across jurisdictions. | Focuses on a single institutional case; qualitative insights only. |

**Stock Markets in Emerging Economies**

The development of stock markets in post-socialist Europe followed non-linear trajectories i.e., irregular and path-dependent progress that varied significantly across countries with respect to timing, pace, and responsiveness to similar reforms, shaped by domestic political economy, macroeconomic conditions, and integration with European financial systems (Claessens et al., 2000, EBRD, 2001, Blommestein, 2000). Existing literature emphasizes that capital market emergence in these settings cannot be fully understood through the lens of developed economies alone; rather, it necessitates a synthesis of institutional sequencing, privatization modalities, and monetary regimes (Kornai, 1992; Claessens et al., 2000; Theodoropoulos and Vojinović, 2005). By the turn of the millennium, notable progress was evident, yet substantial disparities persisted. By 2000, 20 of the 26 transition economies had established formal stock markets, although most remained underdeveloped or dormant. Only Estonia and Hungary achieved capitalization-to-GDP ratios comparable to other emerging markets (World Bank, 2001). Additionally, in 1999, stock market capitalization as a share of GDP ranged from 35% in Estonia and 28% in the Czech Republic, to below 5% in North Macedonia and virtually zero in Armenia, highlighting the pronounced variations in financial sector depth among post-socialist countries (Claessens et al., 2000).

The first wave of capital market formation occurred in the early 1990s as states transitioned from mono-bank structures to two-tier financial systems. While these reforms catalyzed banking sector growth, capital markets remained underdeveloped due to weak institutional trust, persistent macroeconomic volatility, and underdeveloped legal enforcement (Claessens and Djankov, 2000; Stubos and Tsikripis, 2005). During this phase, exchanges primarily served as vehicles for mass privatization, with limited secondary trading. For example, in 2000, North Macedonia's turnover ratio was below 10%, and over 95% of transactions were concentrated in 5% of firms, reflecting extreme illiquidity (Claessens et al., 2001). In a broader context, the average market turnover ratio across all transition economies stood at approximately 30%, while the Baltic states specifically exhibited market turnover ratios of just under 3%, both significantly below the 121% average recorded in comparable emerging markets (World Bank, 2001). Moreover, the vast majority of these exchanges failed to provide reliable alternatives to bank financing or a functioning secondary market for ownership transfers, functions that are critical for post-socialist restructuring (Blommestein, 2000).

As a result of the post-socialist privatization, two main markets emerged:

**Privatization-driven Markets:** Characterized by large numbers of listed firms with minimal trading activity. These markets were found in countries such as Bulgaria, Slovakia, and North Macedonia and were formed through voucher-based privatization, which diluted ownership across atomized stakeholders and fostered low liquidity (Claessens et al., 2000). These systems were designed primarily to expedite ownership transfer rather than enable capital formation, often lacking robust regulatory oversight or institutional investors. Bulgaria's median listed company in 1999, for instance, had annual revenues of just $4 million and 64% insider ownership conditions that discouraged market-based financing (World Bank, 2001, p. 10). Countries such as the Czech Republic and Russia followed similar paths, where mass privatization outpaced the development of market infrastructure and regulation, leading to fragmented, opaque, and often non-transparent trading systems (Blommestein, 2000).

**IPO-led Markets:** These formed through primary listings and targeted foreign capital inflows. In countries like Estonia, Hungary, and Poland, initial public offerings were used to anchor financial development. In Estonia and Latvia, foreign listings comprised over 40% of domestic market cap by 1999, signaling early internationalization (Claessens et al., 2000). Turnover ratios in these countries were substantially higher, 93% in Hungary, 81% in the Czech Republic, and 69% in Poland, compared to under 5% in most Central Asian states (World Bank, 2001, p. 17). These systems were generally supported by more comprehensive legal frameworks, stronger supervision, and institutional arrangements that approached IOSCO standards, particularly in Hungary and Poland (Blommestein, 2000).

This divergence was shaped by the extent of institutional reform, timing of EU accession, and macroeconomic stabilization. Countries pursuing early EU alignment, particularly in Central Europe, developed stronger regulatory regimes and investor protections. Investor protection, measured by indicators such as shareholder rights and enforcement mechanisms, was relatively stronger in Estonia, Hungary, and Poland, facilitating listing and cross-border investment (World Bank, 2001, p. 20). The emphasis on high disclosure standards and centralized, electronic, dematerialized trading further distinguished these markets from their peers (Blommestein, 2000). This translated into more active exchanges and better access to international capital. Moreover, the performance of these exchanges diverged significantly over time. For example, in the mass-privatization markets (Czech Republic, Bulgaria), the number of listed companies declined rapidly after initial peaks, due to illiquidity, tax avoidance concerns, and high disclosure costs. In contrast, IPO-driven systems like those in Hungary and Poland experienced gradual increases in listings and capital raised, albeit from a smaller base (World Bank, 2001).

Another structural issue concerned the dominance of over-the-counter (OTC) and regional trading systems in several mass-privatization markets. In the Czech Republic, for example, only 3% of equity trades occurred on the Prague Stock Exchange by the mid-1990s, with most trading fragmented across multiple systems and weakly regulated platforms (Blommestein, 2000, p. 10). In contrast, Poland implemented dematerialized trading, strong disclosure, and centralized clearance mechanisms, providing a superior foundation for capital formation.

## Comparative Analysis of Baltic States and Western Balkans

The post-socialist transition in Eastern Europe has been marked by dramatic structural transformations as economies moved from central planning to market orientation. Both the Baltic States and the Western Balkans underwent substantial economic restructuring, albeit with divergent paces and institutional trajectories. The early 1990s were characterized by massive GDP contractions typically ranging from 30 to 50 percent and hyperinflation, sometimes exceeding 1000 percent in the Baltics and 3000 percent in the Western Balkans (Kjaergaard and Kjaergaard, 2001; EBRD, 1999). These dislocations induced deindustrialization, fiscal crises, and persistent unemployment, sharply eroding living standards.

However, the Baltic States implemented macroeconomic stabilization programs earlier and more decisively. For instance, by 1995, Estonia, Latvia, and Lithuania had already established macroeconomic stability, underpinned by fiscal consolidation and independent central banks (EBRD, 2001). Inflation dynamics reflected divergent policy responses. Baltic States rapidly curbed inflation through tight monetary policy, reducing inflation from four-digit levels (e.g., Latvia's 958% in 1992) to below 10% by 1997. Western Balkan countries, by contrast, struggled to control inflation, with Yugoslavia experiencing hyperinflation in excess of 3000% in the early 1990s. Only after the adoption of fixed or managed exchange rate regimes did inflation stabilize, particularly in countries like North Macedonia and Croatia.

Fiscal performance also varied. Baltic countries undertook early fiscal consolidation, maintaining budget deficits under 3% of GDP by the late 1990s. Public debt levels remained moderate. In the Western Balkans, fiscal deficits were wider and more persistent, driven by reconstruction needs, public wage pressures, and weaker tax administration. In Serbia, general government fiscal balance was ~ 6.2% of GDP in 2000 (EBRD, 1999, 2001), compared to around ~2.5% in Estonia.

In terms of external orientation, both regions had to pivot away from reliance on former economic unions, the Baltics on the Soviet bloc, the Western Balkans on intra-Yugoslav trade. However, the Baltics reoriented faster, with over 65% of their exports directed toward EU markets by the early 2000s. The region signed numerous free trade agreements, pursued tariff liberalization, and adopted EU regulatory norms earlier and more comprehensively (EBRD, 2001). Western Balkan economies only later moved toward similar trade openness, largely through CEFTA and Stabilization and Association Agreements.

Additionally, the divergence in macroeconomic performance can also be linked to differences in institutional reform and governance quality. The EBRD's Transition Indicators for 2001 show that the Baltic States had progressed further in areas such as large-scale privatization, enterprise restructuring, and competition policy. For example, Estonia and Latvia achieved scores of 4+ (on a 1–4+ scale) for banking reform and enterprise restructuring, compared to lower scores in Western Balkan countries such as Serbia and Bosnia and Herzegovina, which remained around 2–3 (EBRD, 2001). These institutional gaps partly explain the delayed transmission of macro-stabilization to sustained growth in the Western Balkans.

Table 2 provides comparative macroeconomic indicators of the Baltics and Western Balkans during the transition period beginning early 1990s to early 2000s.

**Table 2.** Comparative Macroeconomic Indicators of the Baltic States and Western Balkans during the Transition Period (1990s–early 2000s)

| Indicator | Baltic States | Western Balkans |
|---|---|---|
| Initial GDP Contraction (early 1990s) | Output declined by approximately 30–50% following transition shocks (EBRD, 1999). | Output declined by 40–60%, with severe contraction in conflict-affected economies such as FR Yugoslavia (EBRD, 2001). |

| Indicator | Baltic States | Western Balkans |
|---|---|---|
| EBRD Transition Score: Enterprise Reform | Averaged between 3.5–4.0; reflecting substantial progress in governance, restructuring, and market competition. | Averaged between 2.5–3.0; reflecting slower privatization and persistence of state-owned enterprises. |
| Real GDP Growth (1996–2000, avg.) | Sustained recovery with average annual growth of 4–5%, supported by early reforms and EU-oriented trade. | Growth remained volatile and subdued (~2–3%), hindered by delayed stabilization and post-war reconstruction. |
| Peak Inflation Rate | Hyperinflation exceeding 1,000% in 1992; stabilized to <10% by 1997 through credible monetary policy (EBRD, 1999). | Extreme hyperinflation in Serbia, reaching 3,000%+ in 1993; stabilization delayed until late 1990s (EBRD, 2001). |
| Fiscal Balance (2000) | Budget deficits contained below 3% of GDP; Estonia recorded ~–1.7% of GDP (EBRD, 2001). | Higher deficits: Serbia recorded a general government deficit of –6.2% of GDP in 2000; others ranged –3% to –5% (EBRD, 2001). |
| Export Orientation (EU Share by 2000) | Over 65% of exports directed to EU markets, facilitated by early trade agreements and structural reforms. | Export structures remained regionally concentrated; EU export share below 40% in most cases. |
| Foreign Direct Investment (FDI) | Relatively high FDI inflows; exceeded 5% of GDP in certain years, especially in Estonia and Latvia. | FDI inflows lower, typically <3% of GDP; deterred by regulatory risks and political instability. |
| Trade Integration Instruments | EU Association Agreements signed early; rapid WTO accession; customs and standards harmonization initiated by mid-1990s. | CEFTA and Stabilisation and Association Agreements adopted later; institutional reforms remained uneven. |
| Exchange Rate Regime | Currency boards (Estonia, Lithuania), pegs (Latvia) maintained monetary stability and anchored expectations. | Mixed regimes: euroisation (Montenegro, Kosovo), soft pegs (North Macedonia), managed floats (Serbia). |

Source: EBRD Transition Reports (1996, 1999, 2001)

## Overview of capital markets

Building on the macroeconomic foundations, capital markets offer an additional lens through which to understand structural divergence. The formation of capital markets in both the Baltic States and the Western Balkans was shaped by the legacies of socialist economic systems and the urgency of post-independence institutional reconstruction. As previously discussed, in both regions, stock exchanges were not established to foster capital formation per se, but rather as transitional tools for reallocating ownership under mass privatization schemes (Ginevičius & Tvaronavičiene, 2003). This non-organic emergence resulted in fragile and poorly functioning markets: characterized by illiquidity, weak regulatory oversight, limited secondary trading, and a reliance on state-owned banks for financial intermediation (Claessens et al., 2000).

From this common baseline, the post-1995 evolution of capital markets took markedly divergent paths. This divergence is critical to understanding why the Baltic experience, and specifically the pre-integration structure of the Tallinn, Riga, and Vilnius stock exchanges, offers a valid empirical reference point for the hypothetical consolidation of Western Balkan exchanges. As of the late 1990s, Baltic exchanges offer similarities alike Western Balkan markets in almost every structural dimension: low turnover, small market size, limited foreign participation, and regulatory fragmentation.

Empirical indicators confirm this claim of comparability. In 1998, Estonia's market capitalization stood at just 12% of GDP, increasing steadily to over 32% by 2003. Latvia and Lithuania followed with more modest, yet rising, capitalization ratios 8% and 15.7% respectively by 2003 i.e., pre-integration. Turnover ratios also improved gradually: Estonia surpassed 15%, and Latvia and Lithuania showed moderate but consistent gains. Notably, the number of listed firms in the Baltics declined over time, however due to increased disclosure requirements and listing standards, encouraging quality over quantity. Moreover, technological upgrades, such as Estonia's early adoption of the HEX trading platform were signaling efforts to align with Nordic and EU practices well before formal OMX integration in 2004. In contrast, Western Balkan markets developed at a slower and more fragmented pace. Many exchanges, including those in Bosnia and Herzegovina and Montenegro, either did not exist or lacked consistent data reporting until the early 2000s. Macedonia's and Serbia's exchange, while operational, prior to the 2000s lacked consolidated comprehensive reporting. Slovenia presents a relative exception: its stock market achieved capitalization-to-GDP ratios between 11% and 17% during 1999–2001 and maintained a stable listing base. However, even in the case of Slovenia, liquidity remained low i.e., turnover ratios between 5% and 9% and minimal foreign participation. Ljubljana Stock Exchange bulletins from 2000 and 2001 indicate that despite a rise in the number of listed firms, market concentration remained high and financial intermediation largely bank-driven.

By the early 2000s, the divergence between the two regions had become structural. While the Baltics benefitted from a convergence agenda anchored in EU accession fostering regulatory alignment, investor protections, and enhanced disclosure, the

Western Balkans lagged behind. A major divergence occurred in the early 2000s when the Baltic exchanges pursued regional consolidation. The 2004 merger into the OMX Baltic umbrella, later absorbed by Nasdaq, resulted in a harmonized trading platform, centralized depository infrastructure, and unified listing requirements. This strategic integration significantly lowered transaction costs, enhanced cross-border trading, and increased transparency. For example, Ginevičius & Tvaronavičiene, (2003) find that liquidity improved substantially: the turnover ratio increased to over 30% by 2003 for the more active exchanges, while the market capitalization-to-GDP ratios exceeded 25% in Estonia and Latvia by the mid-2000s. Meanwhile, in the Western Balkans, the persistence of fragmented national exchanges with weak regulatory convergence hindered similar advances.

Taken together, these insights suggest that stock market development in post-socialist Europe was shaped less by organic firm growth and more by privatization sequencing, regulatory convergence, and regional integration frameworks. The limited effectiveness of standalone national exchanges in small economies sets a compelling rationale for exploring regional consolidation such as the Baltic OMX case which this paper empirically investigates through the lens of the Western Balkans. In fact, from a structural perspective, Theodoropoulos and Vojinović (2005) argue that fragmented national exchanges are suboptimal in small economies, given the scale constraints and technology costs. This argument aligns with the broader literature on stock exchange consolidation (Kokkoris and Olivares-Caminal, 2008), which emphasizes liquidity pooling, transaction cost reduction, and investor depth as primary benefits of regional mergers (Click and Plummer, 2005). Fragmentation is not merely an operational inefficiency but a structural impediment to transparency and integration one that reinforces insider systems and "closed-shop" practices often observed in transition economies without robust governance (Blommestein, 2000).

Moreover, the decision to use the Baltic States as reference case in the transfer exercise is rooted in structural similarities from the post-socialist period until the new millennia. To that end we proceed to expand the research in context of financial integration through stock exchange consolidation by employing a synthetic control evaluation of the Baltic Experience Applied to the Western Balkans.

Table 3. Comparative Capital Market Indicators of the Baltic States and Western Balkans

| Indicator | Baltic States | Western Balkans |
| --- | --- | --- |
| Initial Market Purpose | Ownership redistribution via privatization (non-capital raising) | Same |
| Market Capitalization-to-GDP (2003) | Estonia: 32%Latvia: 8%Lithuania: 15.7% | Slovenia: 17%Croatia: 14.7%North Macedonia: 13.6%Others: <5% or no data |

| Indicator | Baltic States | Western Balkans |
| --- | --- | --- |
| Turnover Ratio (2003) | Estonia: >15% Latvia: ~8%Lithuania: ~5.5% | Slovenia: 5–9%Croatia: ~6%Others: negligible or unreported |
| Number of Listed Companies (2003) | Estonia: 14 Latvia: 56Lithuania: 45 | Slovenia: 170+Croatia: 157North Macedonia: 45Others: negligible or unknown |
| Foreign Company Listings | Present but limited | Rare or absent |
| Foreign Investor Participation | Estonia/Latvia: >40% ownership (2003) | Croatia: moderate Serbia/Bosnia: limited, Others: unreported |
| Institutional Investors | Present and increasingly active (e.g., pension funds) | Largely absent or nascent |
| Trading Infrastructure (pre-2004) | Early platform integration (e.g., HEX in Estonia) | Fragmented or underdeveloped systems |
| Regulatory Alignment | Broadly aligned with EU acquis by 2004 | Partial and uneven across jurisdictions |
| Integration Outcome | OMX Baltic merger in 2004 (harmonized trading and clearing) | No integration; national exchanges remain fragmented |

Source: Ginevičius & Tvaronavičiene (2003); EBRD Transition Reports (2000–2004); Ljubljana SE Statistical Reports (2000–2001); National stock exchange yearbooks and author's dataset

3. Methodology and data

3.1 Two-Stage Counterfactual Transmission Design

This paper develops a **two-stage counterfactual transmission design** that links institutional stock-market integration to the dynamics of monetary-policy pass-through. The approach combines the **Synthetic Control Method (SCM)** and **Local Projections (LPs)** within a unified empirical framework, enabling the joint evaluation of both (i) *structural* effects of exchange consolidation on market capitalisation levels, and (ii) *dynamic* effects on the propagation of monetary-policy shocks through the asset-price channel.

The design addresses a key identification problem: since no actual merger of Western Balkan stock exchanges has occurred, standard quasi-experimental tools such as difference-in-differences (DiD) or event studies cannot be applied. These methods require an observable treatment event and pre-intervention parallel trends—conditions not met in a region characterised by institutional discontinuities, shallow markets, and heterogeneous reform trajectories (Campos and Coricelli, 2002; Kovtun et al., 2014).

To overcome this limitation, **Stage 1** constructs a *credible counterfactual of financial integration* via the SCM, while **Stage 2** embeds that counterfactual into an LP framework to trace how monetary shocks would propagate through equity valuations under integrated versus fragmented structures.

To address these identification challenges, this study employs the Synthetic Control Method (SCM), developed by Abadie and Gardeazabal (2003) and later expanded by Abadie, Diamond, and Hainmueller (2010). The SCM is well-suited for estimating policy impacts in settings with a single treated unit real or hypothetical as it constructs a synthetic version of the treated unit from a convex combination of untreated donor units. This synthetic control is selected to closely match the pre-intervention path of the treated unit on key outcome variables and predictors, thereby providing a trajectory of what would have occurred in the 'treated' group i.e., Western Balkans.

The central assumption behind the SCM is that the pre-treatment dynamics of the treated unit can be adequately reproduced by a weighted average of control units. When this condition holds, the difference between actual and synthetic post-treatment outcomes can be interpreted as the causal effect of the intervention.

However, we must acknowledge that the consolidation of the Baltic stock exchanges occurred concurrently with EU accession in 2004, thereby generating a compound treatment effect. As a result, the synthetic control estimates may reflect both the effects of financial market consolidation and the broader institutional alignment associated with EU membership. To address this, the analysis incorporates a comparative SCM using other 2004 EU accession countries without stock market mergers as a partial control group. This involves constructing a parallel SCM for a control group consisting of Central and Eastern European countries such as Slovakia, Slovenia and Poland i.e., states that acceded to the EU in 2004 but did not experience stock exchange mergers. The rationale behind this approach is to attribute observed changes in capital market indicators to EU integration alone, thereby allowing for a comparative estimation of the marginal impact attributable solely to stock exchange consolidation.

This layered SCM design facilitates a difference-in-differences inference strategy. By comparing the synthetic trajectory of the Baltics (which underwent both EU accession and exchange consolidation) with that of the placebo group (EU accession only), the model approximates the incremental contribution of the merger component to observed outcomes such as market liquidity, investor participation, and capital formation.

In parallel to constructing this synthetic counterfactual, the study also examines whether a consolidated market structure would strengthen the monetary-policy transmission mechanism. The rationale derives from empirical literature emphasising that financial integration can amplify the propagation of monetary shocks through equity markets, particularly in small open economies with shallow financial systems (Ehrmann and Fratzscher, 2006; Meier, 2013). To this end, a Local Projections (LP) model à la Jordà (2005) is employed to trace the dynamic responses of market capitalisation to a standardised 100-basis-point policy-rate shock under two structural configurations: (i) fragmented domestic exchanges, and (ii) a counterfactual integrated WB3 exchange, as derived from the SCM simulation (see Section 4).

Hence, the paper employs two complementary empirical strategies:

- **Stage 1:** a Synthetic Control Method simulation assessing the impact of a hypothetical merger on market capitalisation and depth; and

- **Stage 2:** a Local Projections framework evaluating how exchange consolidation alters the elasticity of equity valuations to monetary-policy shocks.

3.1.1 Stage1: Synthetic Control Method

We study three aggregates: the Baltics and EU03 (both "treated" in 2004) and WB3 (treated only in a scenario sense). Primary outcomes are the natural logs of market capitalization in euros $Y_t \in \{lmc\_eur\}$. For each treated aggregate $u \in \{Baltics, EU03 \text{ and } WB3\}$, the synthetic control method (SCM) forms a convex combination of donors that minimizes the pre-treatment distance between the observed path and a synthetic path built from predictors and lagged outcomes.

Let $\hat{Y}_t^{0,\hat{u}}$, denote the synthetic counterfactual and $gap_t^u \equiv Y_t^u - \hat{Y}_t^{0,u}$, the log gap. Pre-treatment windows are 1998–2003 for the Baltics and EU03 and 2001–2008 for WB3. Predictor sets include standard market-development covariates (i.e., firm population in logs, trade openness, log GDP, regulatory quality) and outcome lags over the pre-period. Pre-fit adequacy is summarized by RMSPE, i.e., the root mean squared prediction error of $Y_t^u - \hat{Y}_t^{0,u}$ in the pre period, and by the post/pre RMSPE ratio, i.e., $RMSPE_{pre}$ and $RMSPE_{post.}$

Model adequacy is assessed using the Root Mean Squared Prediction Error (RMSPE), computed as the average pre-treatment deviation between actual and synthetic paths, and by the post/pre RMSPE ratio, which provides a relative measure of fit stability across periods.

Baseline Estimates and Accession Netting

We estimate two standard SCM baselines allowing us to *net out* movements plausibly common to 2004 accession peers.

1. **Baltics baseline (treat = 2004).** We obtain $\hat{Y}_t^{0,Balt}$ and $gap_t^{Balt}$ on the full sample window, with pre-fit diagnostics (i.e., pre-RMSPE and post/pre ratios) and standard placebos.

2. **EU03 baseline (treat = 2004).** We analogously obtain $\hat{Y}_t^{0,EU03}$ and $gap_t^{EU03}$.

To isolate the component specific to **Baltic stock-exchange integration** rather than generic EU accession, we compute the **Baltics-minus-EU03 net gap**:

$$\text{net\_gap}_t \equiv gap_t^{Balt} - gap_t^{EU}, \; t \geq 2004,$$

We index this profile by *relative time* around the 2004 treatment, i.e., r = t - 2004 (so r = 0, 1, 2 …). Then define the transferred profile as the Baltics-EU03 gap in that year:

$$\delta^r = \text{net\_gap}_{(2004 + r)},$$

$\delta^r$ is measured in log point deviations of market capitalisation. We report pre-fit (1998–2003) RMSPE for the net series, the post/pre ratio over primary and extended windows, and placebo-in-space diagnostics for the Baltics experiment, screened by pre-fit quality (i.e., K-screen).

WB3 Baseline and Transfer Mapping

Because WB3 has not undergone an actual exchange merger, we estimate only its no-merger baseline and then apply a transfer mapping as a simulated scenario.

- **WB3 baseline** (pseudo treat = 2009). We estimate $\hat{Y}_t^{0,WB3}$ over 2001-2008 with the same predictor logic and pre-fit checks. The 2009 pseudo date defines WB3 relativetime r = t - 2009.

- **Transfer mapping** (descriptive scenario, not a new causal estimate). We *import only the shape* of the Baltics-specific net profile in relative years, i.e.,

$$Y^{sim}t(s) \equiv \begin{cases} \hat{Y}_t^{0,WB3}, & t \geq 2009, \\ [2pt] \hat{Y}_t^{0,WB3} + s \cdot \delta t - 2009, & t \geq 2009, \end{cases}$$

with transfer share s ∈ {1.0, 0.5} for 100 percent and 50 percent variants. The simulated gap is then $gap^{sim}t(s) = s \cdot \delta t - 2009$ for t ≥ 2009 and zero before. This preserves WB3's own level and trend and adds only the incremental divergence shape observed for the Baltics after removing movements shared with EU03.

Because the Baltics and WB3 pre windows differ (i.e., 1998–2003 versus 2001–2008), we optionally harmonize scales via

$$\text{scale}_{SD} \equiv \frac{sd(gap_t^{WB3} \text{ in WB3 pre})}{sd(net \; gap \; t, \text{in Baltics pre})}, \text{ and use } \delta_r \leftarrow \text{scaleSD} \cdot \delta_r,$$

This ensures proportionality between transferred and native pre-treatment variances.

To contextualise the WB3 scenario without inferring significance, we overlay a placebo-in-space band (5th-95th percentile) computed from good-fit Baltic placebos in relative time. The band is expressed as:

$$Y_t^{band,lo} = Y_t^{0,WB3} + p^5{}_t - 2009 \text{ and } Y_t^{band,lo} Y_t^{0,WB3} + p^{95}{}_t - 2009$$

Hence the WB3 transfer is a *scenario,* i.e., an external-validity thought experiment. We *do not* claim that $gap_t^{sim}(s)$ is a causal estimate for WB3. The exercise asks, if WB3 had experienced a Baltics-specific pattern (i.e., the part not shared with EU03), how would its path have compared to its no-merger baseline?

Donor-set disclosure and disjointness for transfer

To avoid target-in-source circularity, we exclude WB3 and its constituents from the donor pools used to construct the Baltics and EU03 synthetic controls that feed $\delta_r$. In SCM the post-treatment counterfactual is a fixed convex combination of donors; if WB3 entered those donor pools, the $gap_t^{Balt} - gap_t^{EU}$ would mechanically embed WB3's realized path. Transferring $\delta_r$ back to WB3 would therefore re-import its own information, generating **reflection bias** and inflating apparent fit.

This also conflicts with the clean-controls idea of no interference, i.e., donors should be unaffected by the treatment being studied over the evaluation window, and it can distort placebo-in-space diagnostics by letting WB3's own noise leak through donor weights.

We therefore impose a disjoint-donor rule for transfer: when an effect profile is intended for application to a target unit *U, U* and its constituents are excluded from every donor pool used to construct that profile.

3.1.2. Stage 2: Local Projections of Monetary Transmission

To assess whether a hypothetical integration of Western Balkan stock exchanges enhances the transmission of monetary policy via the asset-price channel, we estimate a series of local projections (LPs) à la Jordà (2005). This approach allows the estimation of impulse-response functions (IRFs) to a monetary-policy shock without imposing the dynamic restrictions of VAR frameworks, which is particularly advantageous in settings characterised by short panels and heterogeneous adjustment dynamics such as the Western Balkans.

Whereas the Synthetic Control Method (SCM) provides counterfactual paths for market-capitalisation levels, the LP framework isolates short-to-medium-term responses to exogenous monetary shocks, allowing us to infer how stock exchange integration alters the elasticity of financial valuations to policy impulses. The dynamic response of market capitalisation to monetary-policy shocks is estimated using the following local-projection equation for each horizon *h = 0,1,........24*

$$\Delta y_{t+h} = \beta_h MPshock_{i,t} + \gamma_h + \mu_t + \varepsilon_{i,t+h}, \qquad (1)$$

Where Δ y_{t+h}, denotes the **h-period-ahead cumulative change** in the log of stock-market capitalisation and MPshock$_{i,t}$ is the standardized monetary-policy innovation obtained from the Taylor-rule residuals described below. X$_t$ is a vector of contemporaneous control variables, including inflation, output growth, and changes in official reserves, while $\mu_t$ captures deterministic trends.

The coefficient β$_h$ measures the cumulative response of market capitalisation to a one-standard-deviation monetary-policy tightening. Estimations are performed separately for the **baseline** and **integrated** counterfactuals, and the difference

$$\Delta \beta_h = \beta_h^{int} - \beta_h^{base} \qquad (2)$$

where $\beta_h^{int}$ and $\beta_h^{base}$ denote the estimated responses under the integrated and baseline scenarios, respectively. A negative and statistically significant Δβ$_h$ implies that market capitalisation declines more strongly following a policy tightening under integration, indicating that stock-market integration increases the intensity of transmission through the asset-price channel. The analysis also extends the framework by allowing the effect of monetary shocks to vary with the degree of stock-market integration. This yields an additional coefficient, ghg_hgh , measuring the sensitivity of transmission to integration intensity, and its differential form Δ g$_h$ = $g_h^{int}$ - $g_h^{base}$ , discussed in the main results.

Monetary-policy shocks are derived from an **amended Taylor rule** that explicitly accounts for the hybrid exchange-rate and intervention regimes typical of small open and partially euroised economies. Following Petreski et al. (2025) and Brandão-Marques et al. (2020), we estimate for each country $k$:

$$i_{k,t} - i_{k,t-1} = \alpha_{0,k} + \alpha_{1,k} g_{k,t+1}^F + \alpha_{2,k} \pi_{k,t+1}^F + \alpha_{3,k} g_{k,t-1} + \alpha_{4,k} \pi_{k,t-1} + \alpha_{5,k} f_{k,t-1} + \alpha_{6,k} i_{k,t-1} + u_{k,t} \qquad (3)$$

where $i_{k,t}$ is the domestic lending rate; $g_{k,t}$ and $\pi_{k,t}$ denote real GDP and inflation growth; $g_{k,t+1}^F$ and $\pi_{k,t+1}^F$ represent one-year-ahead forecasts; and $f_{k,t-1}$ is the lagged change in official reserves.

This augmented formulation introduces the reserves term to capture exchange-rate stabilisation motives and balance-of-payments pressures, which standard Taylor rules typically omit. By including lagged real-activity and inflation terms, the specification mitigates simultaneity bias while maintaining a sufficiently parsimonious structure for short samples.

Residuals $u_{k,t}$ represent the unanticipated component of monetary policy, innovations not explained by systematic reactions to observable macroeconomic conditions. These residuals are standardized within each country to account for differing volatility levels, such that a one-unit shock corresponds to a one-standard-deviation policy innovation.

$$z_{k,t} = s_k \frac{u_{k,t} - \bar{u}_k}{\sigma_k}, \qquad s_k = \text{sign}(\text{corr}(\Delta i_{k,t}, u_{k,t})) \qquad (4)$$

where ū$_k$ and $\sigma_k$ denote the mean and standard deviation of $u_{k,t}$ over the estimation window, and $\Delta i_{k,t} = i_{k,t} - i_{k,t-1}$. The s$_k$ sign ensures positive shocks correspond to tightening.

The standardized country shocks are then averaged across Bosnia and Herzegovina, North Macedonia, and Serbia to obtain the aggregate Western Balkans monetary-policy shock series used in the local-projection estimations.

$$\hat{\mathrm{E}}_t^{WB3} = \sum_{k \in (BIH, SRB, MKD)} w_k z_{k,t}, \qquad \sum w_k = 1 \qquad (5)$$

with equal weights w$_k$ = $\frac{1}{3}$ unless otherwise specified. Finally, to maintain comparability across horizons, the aggregate shock is re-standardized:

$$\hat{\mathrm{E}}_t^{WB3} = \frac{\varepsilon_t^{WB3} - \hat{\mathrm{E}}^{WB3}}{\sigma_k(\varepsilon^{WB3})} \qquad (6)$$

The variable $\hat{\mathrm{E}}_t^{WB3}$ is used as the shock regressor in the local-projection estimations, ensuring that all impulse-response coefficients β$_h$ are interpretable as the cumulative effect of a one-standard-deviation tightening in the regional policy stance.

The obtained coefficients offer a direct empirical mapping to the asset-price channel of monetary policy. Contractionary shocks raise the discount rate and lower expected future earnings, thereby reducing equity valuations (Bernanke & Kuttner 2005; Mishkin 2007). Under fragmented exchanges, this adjustment is often attenuated due to shallow market structure, illiquidity and delayed information diffusion. Whereas, under integration, we would expect to observe improved market depth, harmonised disclosure, and a larger investor base enable faster and more complete price realignment, resulting in a more elastic response of market capitalisation to policy changes.

For completeness, the estimated regression coefficients underlying these monetary-policy rules are reported in **Table A1 — Taylor-Rule Estimates by Country (Full Specification) in Appendix 1**, for readers interested in the country-specific policy reaction parameters.

3.2 Data and Variables

The primary outcome variable in our SCM model is the stock market capitalization defined as the total value of shares traded. The first set of predictors **(Table 4)** includes macroeconomic variables that capture the state of the economy. The role **of trade openness** is examined as a key driver of financial integration. According to Pretorius (2002), greater openness facilitates cross-border capital flows and fosters market convergence. Arribas et al. (2006) and Walti (2005) similarly argue that trade liberalization is a precursor to broader globalization processes, where economic openness sets the foundation for financial integration.

In parallel, **GDP** is included to control for country size and level of development. Prior studies such as Edison et al. (2002), Prasad et al. (2003), Vo (2006), and Mishkin (2007) emphasize that richer, more developed economies tend to exhibit higher degrees of international financial integration, making GDP a necessary control in modelling integration effects.

A second group of variables captures macroeconomic characteristics of stock market development. To account for market breadth and investor participation, we consider Beck et al. (2010) and Allen et al. (2011) in employing market size, measured by **the number of listed firms per 10,000 inhabitants**. This variable normalizes market scope by population and enables more accurate cross-country comparisons.

A third set of predictors concerns financial regulation and institutional quality. Vo (2006) shows that institutional constraints and capital controls are key barriers to integration. As such, we incorporate the **Regulatory Quality Index** from the World Bank, which captures the government's ability to formulate and implement sound policies and regulations that permit and promote private sector development.

Finally, a fourth group of variables reflects stock exchange-specific characteristics, which are also found to influence integration. Buttner and Hayo (2011) demonstrate that both absolute and relative market capitalization contribute to equity market co-movements. Tan et al. (2010) similarly highlight how capitalization structure informs financial integration dynamics and to what extent are the benefits of the integration absorbed (Slimane, 2012). Accordingly, market capitalization is included as a direct measure of exchange size.

Table 4. List of covariates employed in model estimation

| Variable | Description / Role in Model | Relevance to Merger Simulation |
|---|---|---|
| **Number of Listed Companies** | Total number of firms listed on the domestic exchange(s) | Indicates market breadth and investor options. Reflects confidence and functionality. |
| **Value Traded (real EUR, 2000 base)** | Total value of shares traded annually, adjusted for inflation and in EUR | Direct measure of exchange liquidity. Evaluates if integration increases actual trading activity. |
| **Market Size (firms per 10k pop)** | Market size adjusted for population | Normalizes breadth across countries. Important for comparing smaller vs. larger economies. |

| Variable | Description / Role in Model | Relevance to Merger Simulation |
|---|---|---|
| GDP (constant prices) | Total real GDP in EUR or USD | Control for economic scale. A key macroeconomic baseline. |
| Trade Openness (% GDP) | (Exports + Imports) / GDP | Captures degree of economic integration with global markets |
| Inflation (CPI, % annual) | Consumer price inflation rate | High inflation may affect investor decisions and stock market stability |
| Regulatory Quality Index | Governance indicator of regulation quality and policy formulation | Stronger institutions are likely to support more successful market integration |

The second empirical stage employs monthly data for 2010–2023, **Table 5**. The analysis focuses on the same WB3 economies but extends the variable coverage to capture high-frequency monetary and real-sector dynamics typical of small open economies.

**Table 5.** List of covariates employed in model estimation

| Variable | Description | Source |
|---|---|---|
| Lending rate | The bank rate that typically reflects the short- and medium-term financing conditions faced by the private sector. It is used both in its level (lagged value) and in its change compared to the previous period. | International Financial Statistics |
| GDP growth | GDP growth rate in real terms. It is used in its lagged value to capture the output gap component of the Taylor rule. | World Economic Outlook |
| GDP growth forecast | One-year-ahead forecast of the GDP growth rate in real terms. Represents policymakers' expectations of future output conditions. | World Economic Outlook |
| Inflation rate | Average consumer-price inflation rate. It is used in its lagged value to account for the backward-looking component of monetary policy reactions. | World Economic Outlook |
| Inflation forecast | One-year-ahead forecast of the average inflation rate. Captures forward-looking behaviour of monetary authorities. | World Economic Outlook |
| Change in reserves | Annual change in official foreign-exchange reserves, including monetary gold, SDRs, and IMF holdings under the control of monetary authorities. It is used in its lagged value to capture external-balance considerations and exchange-rate management motives. | International Financial Statistics |

4. Synthetic Control Results and Merger Simulation

4.1 Baltics — EU Accession and OMX Stock-Exchange Merger

4.1.1 Pre-treatment balance and donor composition

The baseline SCM produces a credible pre-treatment alignment, replicating the Baltics' financial path prior to the 2004 integration. The synthetic Baltics consist of Croatia (about 60 percent), Romania (about 25 percent), and Bulgaria (about 15 percent). This configuration is economically coherent given the shared post-socialist background but differing reform intensities. Croatia contributes institutional maturity and a more advanced financial infrastructure, Romania and Bulgaria introduce variation in liquidity and market scale, i.e., the latter two capture the structural features of larger but less developed exchanges. The donor mix therefore provides a balanced counterfactual that mirrors the Baltics' pre-accession reform environment without duplicating its EU-oriented trajectory.

With only three donors, weight concentration is unavoidable, i.e., Croatia's share is large. This can raise concerns about single-donor dependence. Two points mitigate this: first, the purpose is to approximate the treated unit's pre-trend in the outcome, not to replicate every predictor; second, the fit is evaluated transparently via the pre-period RMSPE and lag matching. As long as the lag structure is closely reproduced and diagnostics are reported, high weights on the most similar donor are standard in small, regionally coherent pools.

Table 6 summarises the predictor balance for 1998–2003. The synthetic unit tracks the treated series closely across the lagged outcome (log market capitalisation), with deviations narrowing from 0.32 to 0.02 log points. This indicates that the pre-treatment trend is well replicated. Some differences appear in macro predictors: the consumer price index is substantially higher in the synthetic (around 7.8 percent) than in the Baltics (around 2 percent), reflecting donors' delayed monetary stabilisation and weaker inflation-anchoring frameworks in the early 2000s. Trade openness (trade-to-GDP) is roughly 20 percentage points lower, which corresponds to the Baltics' earlier and deeper reorientation to EU markets, i.e., over 65 percent of exports already headed to the EU by 2000 compared to lower shares in the donors. These gaps are consistent with broader regional differences: the Baltics entered the 2000s with stronger institutions, tighter fiscal discipline, and higher regulatory convergence, while the donors still faced residual inefficiencies from slower privatisation and weaker governance (EBRD, 2001).

Table 6. Predictor balance and donor weights, Baltics

| Metric | Treated | Synthetic | Difference |
|---|---|---|---|
| *cpi* | 1,80 | 7,77 | -5,97 |
| *lmc_eur(1998)* | 6,95 | 6,83 | -0,12 |
| *lmc_eur(2000)* | 7,96 | 7,75 | 0,12 |
| *lmc_eur(2001)* | 8,00 | 7,76 | 0,23 |
| *lmc_eur(2003)* | 8,84 | 8,56 | 0,28 |
| *ln_firm_pop* | -2,36 | -2,24 | -0,12 |
| *log_gdp* | 10,93 | 10,56 | 0,31 |
| *lturnover_eur* | 6,74 | 6,49 | 0,25 |
| *reg_q* | 0,96 | 0,60 | 0,31 |
| *trade_gdp* | 100,57 | 80,13 | 20,45 |

Source: Author's calculations

The CPI and openness gaps could bias levels of market capitalisation independently of the treatment, i.e., tighter anchors and greater external exposure are typically pro-valuation. We address this by prioritising the match on outcome lags, which carry the strongest identifying content in SCM when predictor sets are short and samples are volatile. In addition, later robustness checks (placebo-in-space, leave-one-out) are used to see whether the post-2004 divergence is unique to the Baltics or reproducible among donors with these higher-CPI, lower-openness profiles.

Other variables such as log GDP, turnover, firm density, and regulatory quality show moderate gaps of about 0.25 to 0.31 log units. These are consistent with regional heterogeneity noted in transition reports, i.e., the Baltics entered the 2000s with stronger institutions, tighter fiscal frameworks, and more advanced disclosure, while donors still faced residual inefficiencies from slower privatisation and weaker governance. In the context of SCM, such deviations are acceptable when the lag structure is tight, since outcome lags receive the greatest weight in the optimisation and the objective is to recover the counterfactual path of the dependent variable rather than exact equality on every covariate.

In practical interpretation, the synthetic Baltics can be viewed as a plausible delayed-integration scenario, i.e., a counterfactual representing how the region's capital markets might have evolved under slower institutional reform and later integration. The pre-RMSPE of 0.33 confirms a solid pre-fit, indicating that residual variation is minimal and the counterfactual provides a solid basis for evaluating post-accession divergence.

4.1.2 Outcome dynamics

**Figure 1** plots the observed and synthetic paths of log market capitalisation. The pre-2004 trajectories overlap closely, reflecting the excellent baseline fit (pre-RMSPE ≈ 0.33). A visible divergence appears after the 2004 accession and OMX integration: the treated series rises above the synthetic in the early post-accession years and maintains a higher level through 2023. This suggests that the joint policy shock produced a persistent upward level effect in market capitalisation relative to the counterfactual.

**Figure 1.** Actual vs. synthetic log market capitalization, 1998–2023

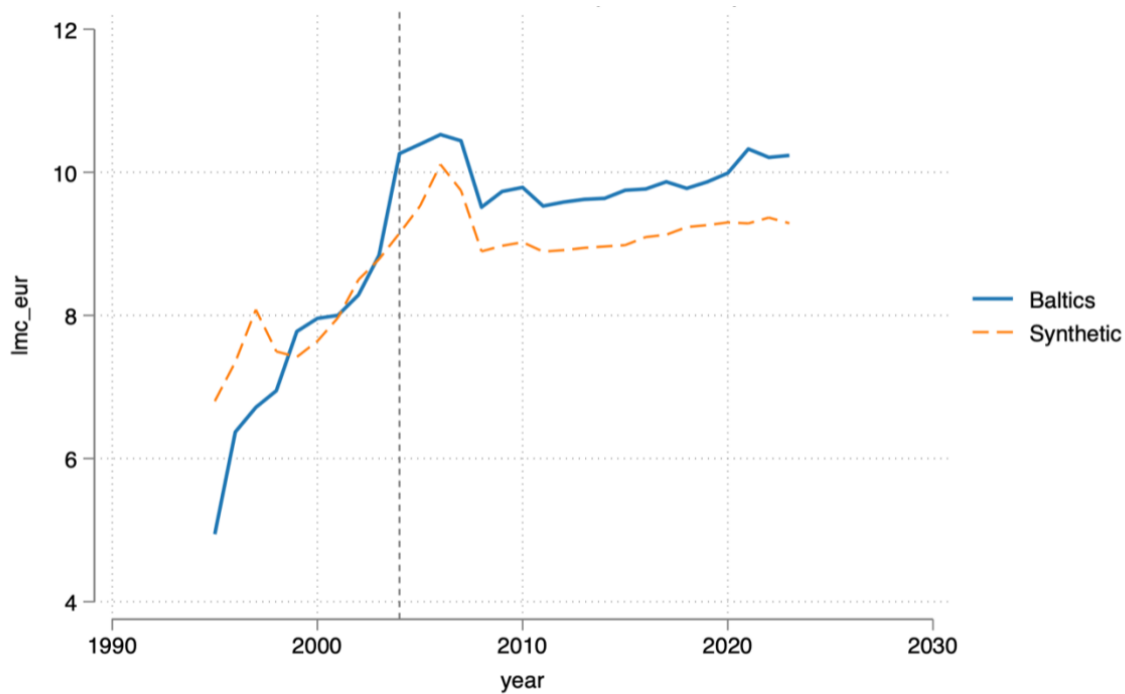

Source: Author's Calculations

**Figure 2** confirms that the treatment effect emerges immediately after 2004, with the gap peaking around 0.3–0.4 log points (roughly 30–40 percent in levels) by 2006 and stabilising thereafter. This trajectory is consistent with the notion of a structural realignment rather than a temporary boom: the merger may have initially reduced fragmentation and increased liquidity, while EU membership improved investor confidence and institutional credibility. Comparable post-integration adjustments are documented in other regional mergers (Nielsson, 2009).

**Figure 2.** Gap (treated – synthetic), 1998–2023

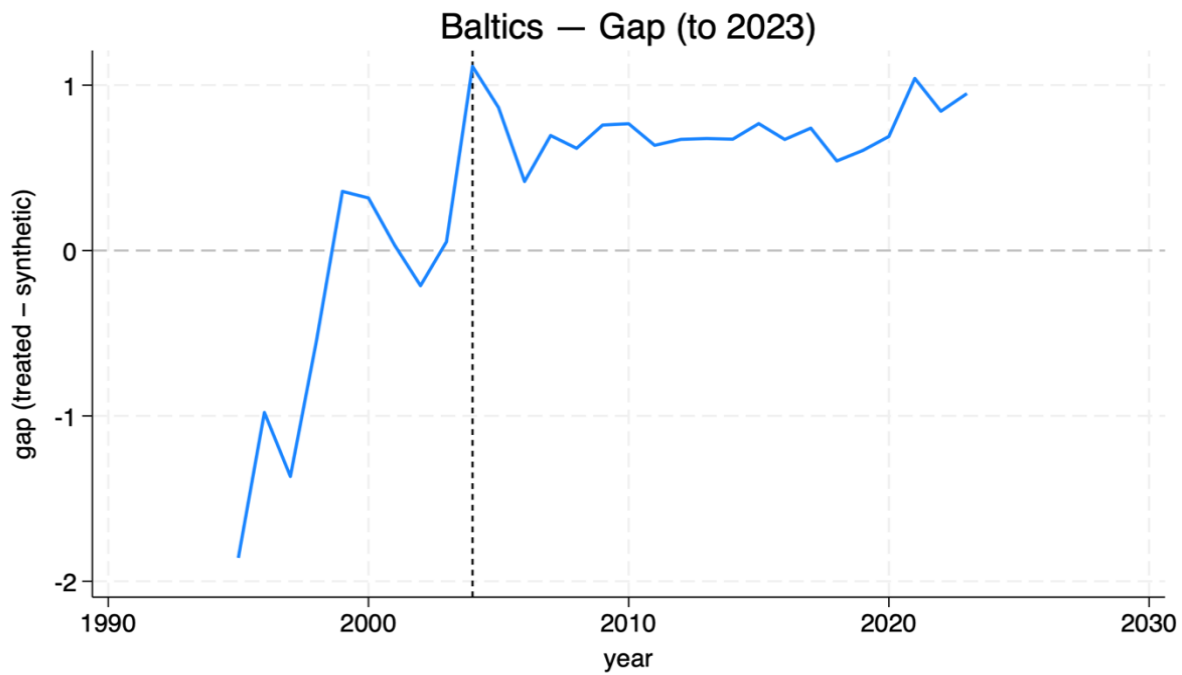

Source: Author's Calculations

4.1.3 RMSPE and magnitude of divergence

Quantitatively, the pre-RMSPE of 0.33 (**Table 7**) indicates a solid pre-treatment alignment. The post/pre RMSPE ratio increases to 2.62 for 2004–2006 and 2,26 for 2004–2023, showing that model error roughly doubled in the short-run window and remained moderately elevated over the full horizon. Ratios of this scale suggest an **economically relevant but not conclusive** deviation, particularly given the small donor pool and macro-financial setting. Overall, the results imply that following EU accession and the regional stock-market consolidation, the Baltics' combined market capitalisation expanded by roughly twice the amount that would have been expected under a no-accession, no-merger counterfactual represented by the synthetic control. However, since the ratio lies in the lower range i.e., around two rather than four or higher, the evidence should be viewed as suggestive rather than conclusive.

**Table 7.** Pre-and post-intervention RMSPE values and ratios, Baltics

| pre_RMSPE | post_RMSPE_0406 | post/pre_0406 | post_RMSPE_0423 | post/pre_0423 |
|---|---|---|---|---|
| 0,3329 | 0,8728 | 2,621 | 0,7548 | 2,2672 |

Source: Author's Calculations

4.1.4 Interpretation and discussion

Overall, the baseline SCM indicates a modest yet persistent rise in the Baltics' market capitalisation following EU accession and the OMX merger. The close pre-treatment alignment lends credibility to the counterfactual, while post-2004 divergence though moderate in scale, is consistent across horizons and within the range typically regarded as suggestive in small-N SCM applications (Billmeier & Nannicini, 2013).

However, several limitations qualify this finding. The short calibration period (1998–2003) constrains matching precision, and minor predictor imbalances on inflation and trade openness suggest that the synthetic represents a slightly less open, more inflation-prone counterfactual. These caveats imply that while the effect is directionally robust, it should be interpreted as indicative rather than conclusive.

Taken together, the results point to an economically meaningful uplift in market valuation i.e., likely reflecting improved investor confidence, institutional credibility, and liquidity conditions under the merger, followed by gradual normalisation. The next subsection tests the stability and distinctiveness of this pattern through placebo and leave-one-out robustness checks.

4.2. EU Accession without Exchange Consolidation

Following the Baltic treatment in 2004, this subsection analyses an aggregate comprising Slovenia, Slovakia, and Poland; hereafter EU03, three of the ten economies that joined the European Union in the 2004 enlargement. Prior studies find that EU accession typically stimulates financial development through improved institutional quality, market integration, and investor confidence (Campos et al., 2019). Accordingly, EU entry in 2004 is modelled as the policy intervention, with the Synthetic Control Method (SCM) used to estimate its effect on log market capitalisation, a proxy for financial depth. We interpret our estimand as a *acceleration* effect: because key donors (RO/BG in 2007; HR in 2013) also acceded later, the synthetic should be viewed as a *delayed-integration benchmark* rather than a pure non-accession counterfactual.

4.2.1 Pre-treatment balance and donor composition

The baseline SCM for EU03 delivers a **robust** pre-treatment fit, effectively reproducing the treated group's financial trajectory prior to EU accession in 2004. The synthetic composite is primarily built from **Croatia (≈55%), Romania (≈27%), and Bulgaria (≈18%),** reflecting geographic proximity and similar post-transition reform paths. This composition is economically plausible in light of transition-era differences in market depth and institutional quality (e.g., EBRD, 2003; Berglöf & Bolton, 2002), with Croatia providing higher institutional/regulatory alignment and Romania/Bulgaria supplying scale/liquidity variation.

Table 8. Predictor balance and donor weights, EU03

| Metric | Treated | Synthetic | Difference |
|---|---|---|---|
| lmc_eur(1998) | 8,02 | 7,40 | 0,63 |
| lmc_eur(2000) | 8,05 | 7,55 | 0,51 |
| lmc_eur(2001) | 8,18 | 7,79 | 0,40 |
| lmc_eur(2002) | 8,51 | 8,23 | 0,28 |
| lmc_eur(2003) | 8,58 | 8,35 | 0,24 |
| ln_firm_pop | -2,72 | -2,27 | -0,46 |
| log_gdp | 12,52 | 11,61 | 0,91 |
| lturnover_eur | 7,39 | 6,33 | 1,05 |
| reg_q | 0,68 | 0,11 | 0,56 |
| trade_gdp | 75,04 | 70,02 | 5,02 |

Source: Author's calculations

As reported in **Table 8**, deviations in log market capitalisation narrow steadily from 0.63 (1998) to 0.24 (2003), confirming the synthetic's ability to track the treated unit's upward path. Residual predictor gaps, in particular lower log GDP (11.6 vs 12.5) and regulatory quality (0.11 vs 0.68) imply that the counterfactual represents a smaller, less institutionally developed market. In practice, this means that the treated economies entered accession from a stronger macro-institutional base, making their post-2004 improvement appear comparatively amplified.

Turnover differences (≈1.05 log points) signal deeper trading activity in the treated group, while firm-density gaps (≈0.46 log units) remain within the acceptable imbalance thresholds for SCM (Kaul et al., 2018). Trade openness aligns closely (75 % vs 70 %), strengthening the plausibility of the donor mix. The pre-RMSPE of 0.23 indicates tight pre-fit precision, suggesting that residual variance is minimal and the constructed counterfactual is credible.

Overall, the donor composition yields a coherent benchmark; one that approximates a delayed-accession scenario within the same regional and institutional context against which the post-treatment divergence of EU03 can be interpreted.

4.2.2 Outcome Dynamics

**Figure 3.** Actual vs. Synthetic log market capitalization, 1998–2023, EU03

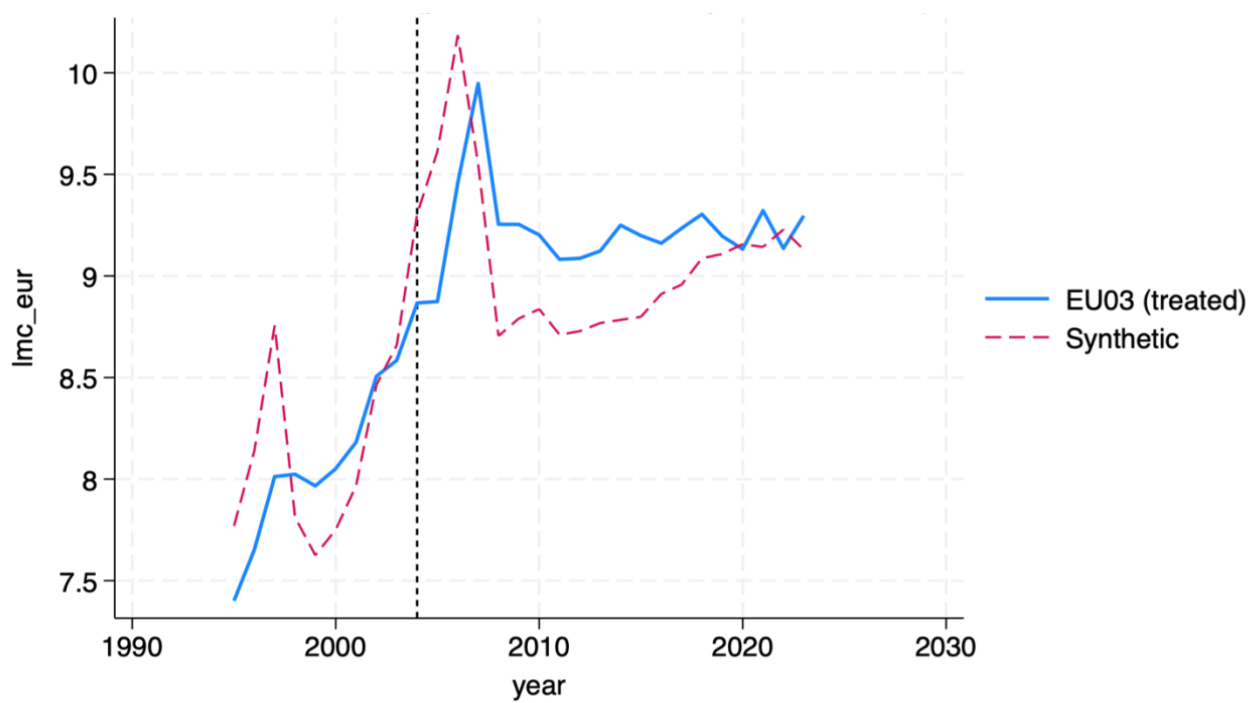

Source: Author's Calculations

**Figure 3** plots the evolution of log market capitalisation for EU03 and its synthetic counterpart. The pre-2004 period exhibits a near-perfect co-movement, confirming that the synthetic closely replicates the treated path prior to accession. After 2004, EU03's market capitalisation rises above its counterfactual, reflecting an accession-related uplift driven by capital inflows and expectations of regulatory convergence.

The synthetic path, reflecting late-accession donors, briefly surpasses the treated line around 2007—coinciding with the regional investment boom preceding the global financial crisis, however the recovery of EU03 is faster and more stable in the subsequent period. Through the 2010s, the synthetic gradually converges as donor economies themselves join the EU, implying harmonisation through shared integration rather than mean reversion. Evidence from the euro area supports the notion that the adoption of common macro-prudential rules under the EU and EMU frameworks strengthened financial interlinkages and advanced stock market integration via macroeconomic harmonisation and financial development (Hardouvelis, 2007; Kim et al., 2005).

**Figure 4.** Gap (treated – synthetic), 1998–2023

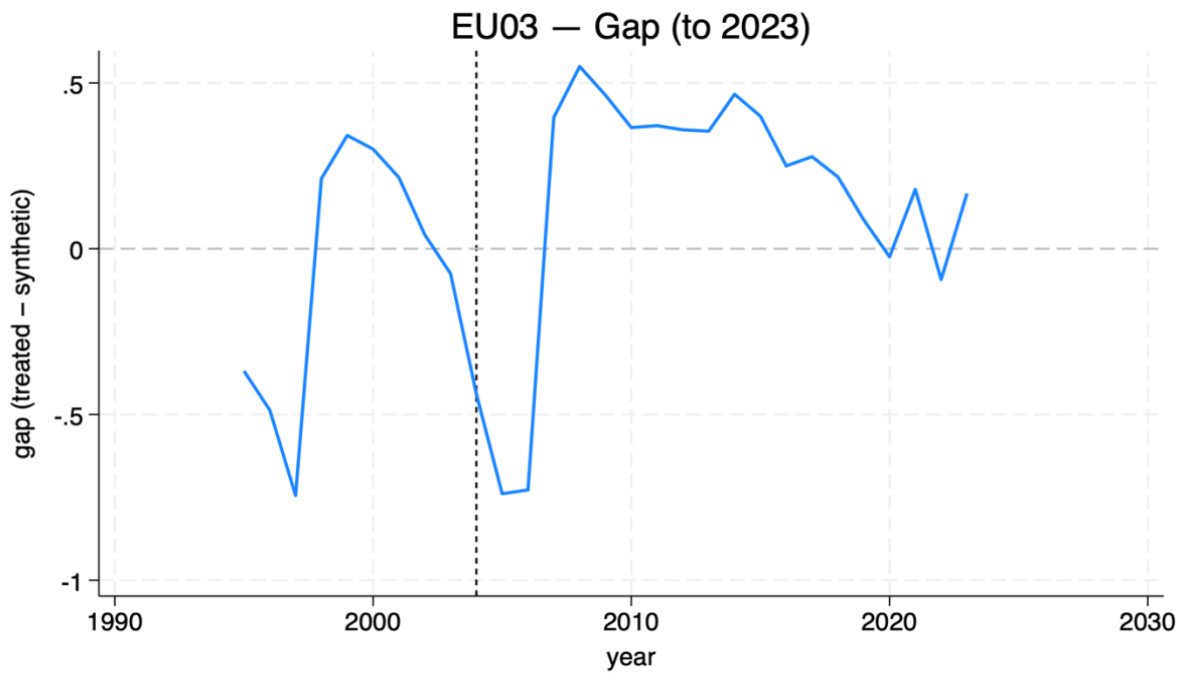

Source: Author's Calculations

The difference peaks near 0.6–0.7 log points by 2006 (≈60–70% in levels) and later stabilises near 0.25–0.3 in the 2010s. i.e., by 2006, EU03's market capitalisation is roughly two-thirds higher than the synthetic as shown in **Figure 4**; in the longer run the gap settles to about one quarter to one third. We view the later convergence as consistent with single-market integration and regulatory alignment, rather than a fading of the accession effect. By the post-2010 period, the treated and synthetic paths evolve in tandem, reflecting the normalisation of financial structures within a common European framework.

4.2.3 RMSPE and Magnitude of Divergence

The RMSPE diagnostics confirm the credibility of this pattern and help gauge the strength of the estimated divergence. The pre-RMSPE of 0.23 demonstrates excellent pre-treatment alignment, well within the < 0.35 threshold commonly used for small-sample SCMs.

Post-accession, model error rises moderately: the post/pre RMSPE ratio equals 2.8 for 2004–2006 and 1.7 for 2004–2023. These values fall squarely within the range regarded as economically meaningful yet methodologically stable (Billmeier & Nannicini, 2013; Becker & Klößner, 2018). Hence, the divergence captured in Figure 4 reflects a genuine policy effect rather than noise or poor pre-fit. Taken together, the RMSPE results corroborate the visual evidence: EU03 experienced a pronounced short-run accession premium that gradually settled into a higher steady state of market capitalisation.

4.2.4 Interpretation and Discussion

The SCM evidence for EU03 indicates a moderate yet durable divergence in market capitalisation after EU entry, mirroring the Baltics' experience in direction though not in magnitude. The short-run spike (ratios ≈ 2.8) denotes an initial acceleration driven by improved confidence and cross-border capital mobility, while the long-run moderation (≈ 1.7) suggests stabilisation at a higher equilibrium of financial depth rather than a reversion.

However, several caveats apply. The short pre-treatment period (1998–2003) limits the precision of the baseline alignment, and the small-N donor base, three countries, two of which later joined the EU reduces the model's statistical power. Hence, the counterfactual should be interpreted as a delayed-integration trajectory rather than a pure "non-accession" path. In this sense, the treatment captures absolute acceleration up to 2007 and relative effect afterwards.

In applied terms, the results imply that EU03's capital markets experienced an early expansion that would likely have been slower without membership, followed by convergence through structural integration as markets matured. The effect resembles a sustained EU membership premium i.e., strong initial inflows and valuation gains that were subsequently consolidated within an integrated financial framework.

Overall, the evidence points to a persistent and economically meaningful increase in market capitalisation associated with accession, absorbed into a longer-run process of structural financial deepening. Given the limited donor pool, overlapping accession among comparators, and brief pre-window, the results should be viewed as indicative yet directionally robust.

4.3 Consolidation Premium (NET)

The Baltics–EU03 net gap isolates the component of the post-2004 divergence that cannot be attributed to EU accession alone but instead likely reflects the structural consolidation of the Baltic exchanges under the OMX framework. Both aggregates acceded to the European Union at approximately the same time; however, only the Baltics underwent a coordinated exchange merger that unified trading, clearing, and listing standards across Estonia, Latvia, and Lithuania. The differential trajectory therefore captures the incremental contribution of this merger.

**Figure 5** shows that immediately after 2004, the net series turns markedly positive, implying that the Baltics' stock-market capitalisation expanded relative to their synthetic path by more than the EU03 aggregate did relative to its own counterfactual. This pattern suggests that the OMX merger produced an additional valuation uplift beyond the accession premium, in line with the theoretical and empirical literature that associates stock-exchange consolidation with enhanced liquidity, reduced transaction costs, and higher investor confidence. Studies on earlier regional integrations such as Euronext or BME document similar mechanisms: structural mergers consolidate fragmented trading environments, improve market depth, and generate operational efficiencies through standardised trading platforms (Pagano et al., 2002; Floreani & Polato, 2010; Dorodnykh, 2013). Empirical assessments further show that such mergers tend to reduce explicit and implicit trading costs, broaden investor participation, and foster informational efficiency (Pagano & Padilla, 2005; Nielsson, 2009).

**Figure 5.** Net Gap Baltics vs. EU03

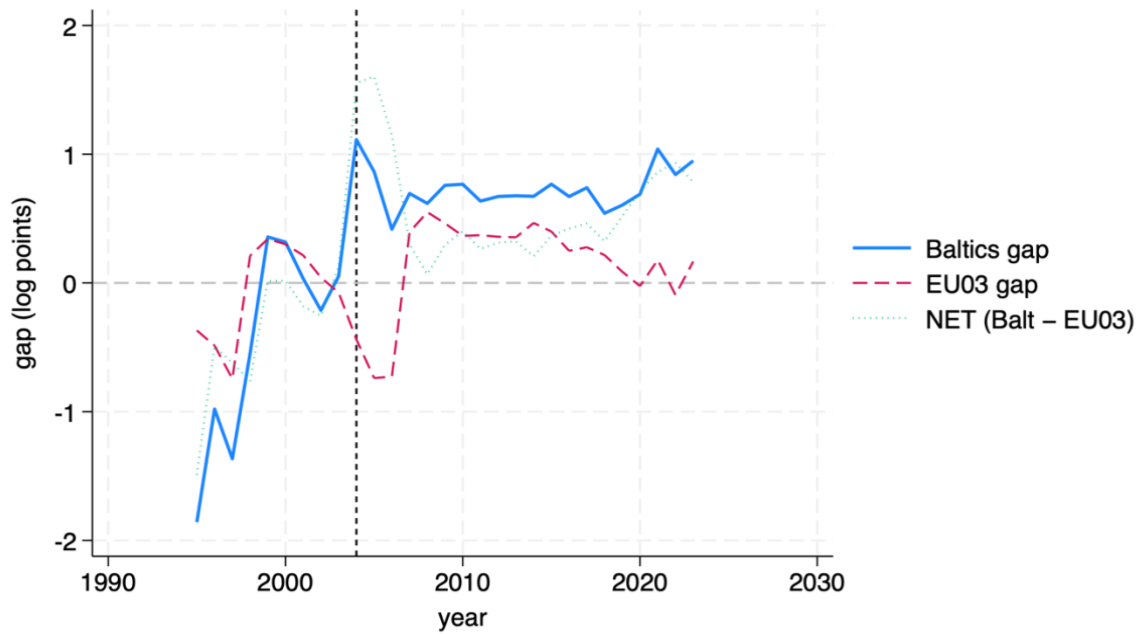

Source: Author's Calculations

Hence, the Baltics' early participation in the OMX network likely accelerated market integration by pooling liquidity and modernising post-trade infrastructure at a time when EU03 exchanges remained institutionally fragmented. Yet, by the mid 2010s, EU03 economies (Poland, Slovakia and Slovenia) reduce the cap partly as post-crisis recovery and partly due to integrated EU frameworks.

In applied interpretation, the results highlight two aspects. First, the pronounced divergence in the early post-2004 years reflects the first-mover advantage of the OMX merger, which delivered a unified market infrastructure while regional peers were still adapting to EU-level norms. Second, the gradual narrowing thereafter reflects policy diffusion and institutional catch-up, as the wider Central- and Eastern-European markets converged within the single-market framework. Hence, the estimated net effect does not capture transient accession 'benefit', but rather the sustained structural payoff of the exchange level-integration.

## 4.4. Simulated Merger Effects: Western Balkans, Transfer Exercise

### 4.4.1 Key Results

The simulated merger exercises provide a structured assessment of how stock-exchange consolidation might influence market-capitalisation in the Western Balkans. To do this, we transfer the post-merger effect observed in the Baltic exchanges after their integration within the OMX system to the Western Balkan Three (WB3), namely Serbia, Bosnia and Herzegovina, and North Macedonia. The simulations are designed to capture both the direct valuation impact (i.e., the difference in market capitalisation between the treated and synthetic series) and the broader structural effects that such consolidation may entail. The latter are inferred from the persistence and trajectory of the simulated uplift (Figure 6), which, if sustained, would be consistent with improvements in liquidity, transparency, and market efficiency typically associated with institutional integration (Pagano and Padilla, 2005; Dorodnykh, 2013). To ensure comparability, the transferred Baltic effect is standardised by the pre-treatment standard deviation of WB3's own market volatility, so that the simulated uplift reflects proportional rather than absolute variation.

To ensure comparability, the transferred Baltic effect is standardised by WB3's pre-treatment standard deviation of market volatility, ensuring that the simulated uplift reflects proportional rather than absolute variation. The results reveal a moderate but persistent divergence between the simulated "with-merger" and baseline synthetic paths, peaking in the early post-simulation period and remaining positive thereafter. This suggests that a hypothetical WB3 consolidation would likely have produced a sustained, credibility-driven increase in market capitalisation rather than a short-lived speculative surge.

**Figure 6.** WB3, Simulated 'with merger' (transfer of Baltics NET)

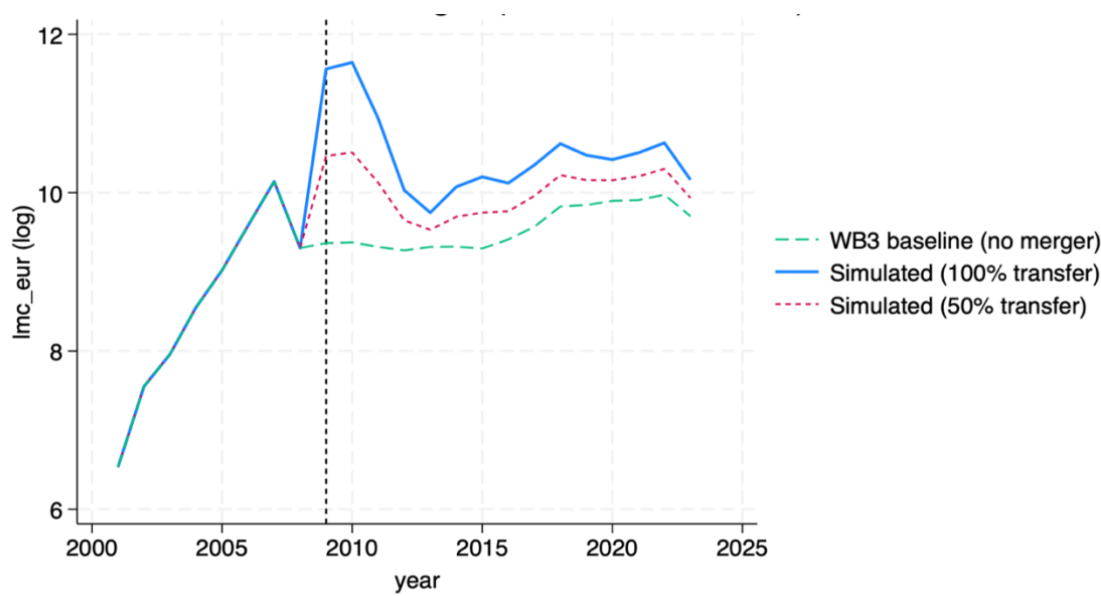

Source: Author's Calculations

4.4.2 Discussion of simulated effects

The results suggest that a WB3 stock-exchange merger would have acted primarily as an institutional lift-off mechanism that could have provided a credibility signal within the region's fragmented financial systems. Evidence from prior consolidations indicates that exchange integration can serve as a policy anchor for broader institutional modernisation, particularly when national markets suffer from limited liquidity and investor confidence (Pagano and Padilla, 2005; Dorodnykh, 2013). The simulated uplift observed for WB3 may therefore be interpreted as a credibility-enhancing response, i.e., a revaluation consistent with improved expectations regarding market continuity, policy commitment, and regulatory coherence. In this sense, the merger would not only have acted as a mechanical lift-off for market capitalisation, but also as a driver of perceived stability. From an institutional standpoint, such integration would have reduced informational and operational fragmentation between the three domestic exchanges. The Baltic experience after their inclusion in the OMX system provides a reference point for this mechanism: the unification of trading standards and clearing procedures contributed to improved informational efficiency and facilitated cross-border participation (Dorodnykh, 2013; Jazepčikaite, 2008).

Such a lift-off mechanism operates not through immediate expansion in market volume but through reinforcement of institutional expectations. In small, transition economies, credibility and coordination often precede liquidity deepening (Hasan et al., 2010; Pagano et al., 2001). Once an integrated exchange is perceived as stable and reform-oriented, it can attract broader investor participation, which in turn strengthens market depth. Hence, the initial post-merger uplift reflected in the simulation can be viewed as the early stage of a credibility cycle, whereby institutional reform itself induces renewed investor confidence and incremental valuation growth.

The persistence and magnitude of the simulated uplift, however, depend critically on the region's **absorptive capacity**, i.e., its ability to internalise and sustain the potential benefits of integration given the structural and institutional characteristics. The literature on exchange mergers highlights that consolidation effects are rarely uniform and that the extent of realised gains depends on pre-existing levels of liquidity, regulatory quality, and investor participation (Dorodnykh, E, 2014; Polato and Floreani, 2010). Smaller exchanges tend to exhibit lower absorption capacity because their institutional frameworks are weaker and their trading bases narrower, which limits the transmission of integration-related efficiency gains. In this respect, the WB3 economies share several features with earlier low-capacity cases: thin trading volumes, a small number of listed firms, low turnover ratios, and a limited domestic investor base. These features constrain the mechanism through which structural integration translates into tangible liquidity improvements.

Yet, Slimane (2012) finds that this limitation is inherently paradoxical, as documented by the case of Portugal within the Euronext merger. Despite its smaller market size, Portugal experienced the strongest volatility-reducing effects, suggesting that structural inefficiencies can create greater scope for adjustment. Hence, low absorptive capacity, while constraining in the short run, may amplify the relative long-term benefits of consolidation once reform and integration advance jointly.

The absorptive-capacity constraint can also be understood in terms of institutional readiness. Integration requires not only harmonised trading technology and regulation but also the administrative ability to coordinate disclosure standards, clearing arrangements, and supervisory practices. Without these complementary capacities, the sustained structural effects of the integration may be hindered. A similar pattern is observable in WB3, where low turnover and investor risk aversion continue to limit the capacity to internalise external efficiency gains.

Even when interpreted through the lens of limited absorption, the simulated merger effects appear economically meaningful, though clearly bounded. The results indicate that exchange integration could have served as a catalyst for gradual financial deepening in the WB3, expanding the capitalisation base without necessarily achieving rapid convergence toward the more developed European markets. This pattern reflects the incremental nature of institutional change in small transition economies, where progress occurs through credibility gains and liquidity accumulation rather than through discrete structural shifts (EBRD, 2006; Hasan et al., 2010). The persistent divergence between the simulated and synthetic trajectories throughout most of the post-2009 period suggests that consolidation would likely have yielded a sustained, though measurable yet contained, increase, consistent with improvements in perceived market stability and transparency.

From a policy standpoint, such an outcome would carry practical significance. In financial systems dominated by bank intermediation, even modest increases in market capitalisation can enhance the role of equity markets in resource allocation and risk sharing (Mishkin, 2007). These effects do not imply a structural transformation comparable to that of the Baltics but still represent an incremental improvement in the financial landscape of the region. In this vein, the simulated merger captures a form of institutional convergence, i.e., a gradual shift toward a more credible and integrated market environment that remains conditional on policy and institutional reform.

Hence we consider that even modest integration could yield meaningful improvements in credibility and capital-market depth, provided it is accompanied by complementary reforms in governance, investor protection, and financial infrastructure. Without these, the positive trajectory may taper, reinforcing that a WB3 merger could contribute to measurable progress in market development within the limits of structural capacity and readiness.

To reinforce the credibility of these findings, the analysis further employs a placebo-adjusted simulation that positions WB3 results within a broader inferential framework. The placebo-in-space procedure (Figure 7) tests whether the simulated uplift in market capitalisation exceeds the range of variation that would typically arise among comparable economies without a merger. In practice, the placebo band represents the dispersion of pseudo-treatment effects across countries that were not subject to integration. When the WB3 "with-merger" trajectory consistently remains above this range, as observed between roughly 2010 and 2018, the outcome suggests that the simulated uplift is unlikely to reflect random fluctuations or measurement noise. Rather, it points to a systematic structural effect that is statistically distinct from placebo outcomes and therefore economically meaningful.

**Figure 7.** Simulated paths with placebo bands

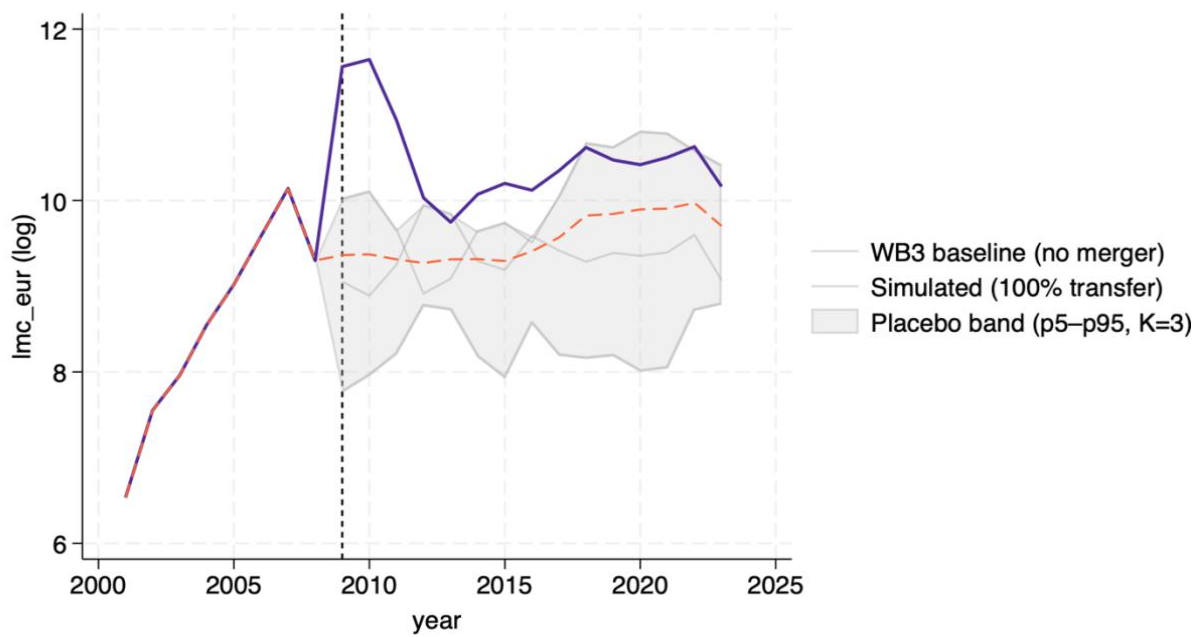

Source: Author's Calculations

Building on these findings, the next section explores whether the same institutional integration would enhance the effectiveness of monetary policy, .particularly through the asset-price channel. Hence we seek to answer what integration does to market depth to how it changes the way policy transmits through those markets.

5. Monetary Policy Transmission

5.1. Key Results

Building on the simulated-merger exercise, this section examines whether the market deepening implied by exchange integration strengthens the transmission of monetary policy through the asset-price channel. The analysis estimates local projections of market-capitalisation responses to a standardised 100-basis-point policy-rate shock under two settings: *(i)* fragmented domestic exchanges and *(ii)* a counterfactual integrated WB3 exchange.

Market capitalisation is used as a proxy for market depth and valuation capacity, capturing the extent to which equity markets react to macro-financial shocks. The shock is normalised to ensure an equivalent initial impact across the two scenarios, so that any divergence in responses reflects differences in structural characteristics rather than in the magnitude of the disturbance.

**Figure 8** illustrates that monetary tightening leads to a modest short-term fall in market capitalisation under fragmentation, but the effect is more pronounced and persistent once stock exchanges are integrated. The divergence becomes visible within the first two years and peaks around horizons 3–4, when the integrated trajectory lies roughly 0.4–0.6 log points below the baseline. Both paths subsequently converge back toward equilibrium, consistent with the transitory nature of the underlying shock.

**Figure 8.** Impulse response of WB3 market capitalization to a 1-SD monetary policy shock under baseline (fragmented) and integrated-exchange scenarios

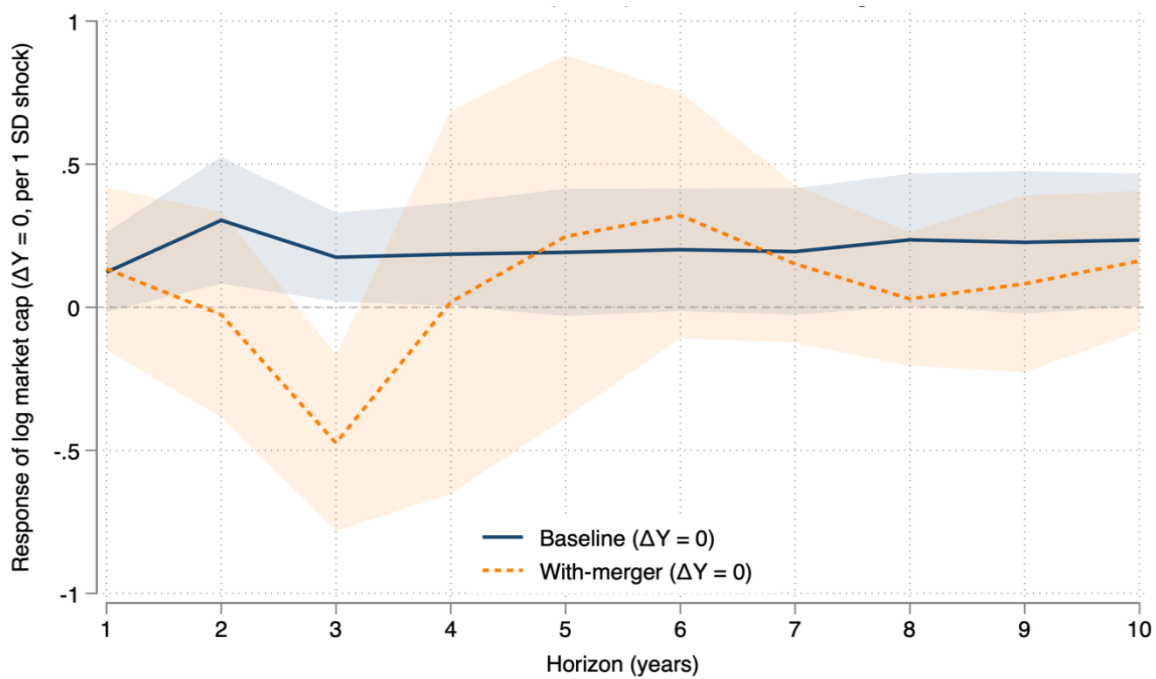

Source: Author's Calculations

*Note: Shaded areas denote 90% confidence intervals.*

The results align with the idea that integrated markets adjust more effectively to policy shocks. Under a unified institutional framework, valuation responses unfold more swiftly and persist over longer horizons, suggesting that integration strengthens monetary transmission without changing its overall direction.

To visualize the marginal benefit of integration, **Figure 10** traces how the sensitivity of market capitalization to monetary-policy shocks changes as stock exchanges become more integrated. The coefficient $\Delta g_h$ declines steadily for roughly three years after consolidation, falling by about 1.5 log-points per one-standard-deviation increase in integration intensity. This trajectory suggests that most liquidity and information-efficiency gains are realized early, as unification channels capital toward deeper and more responsive markets.

**Figure 9.** Merger effect on integration sensitivity ($\Delta g_h$), 90 % CI

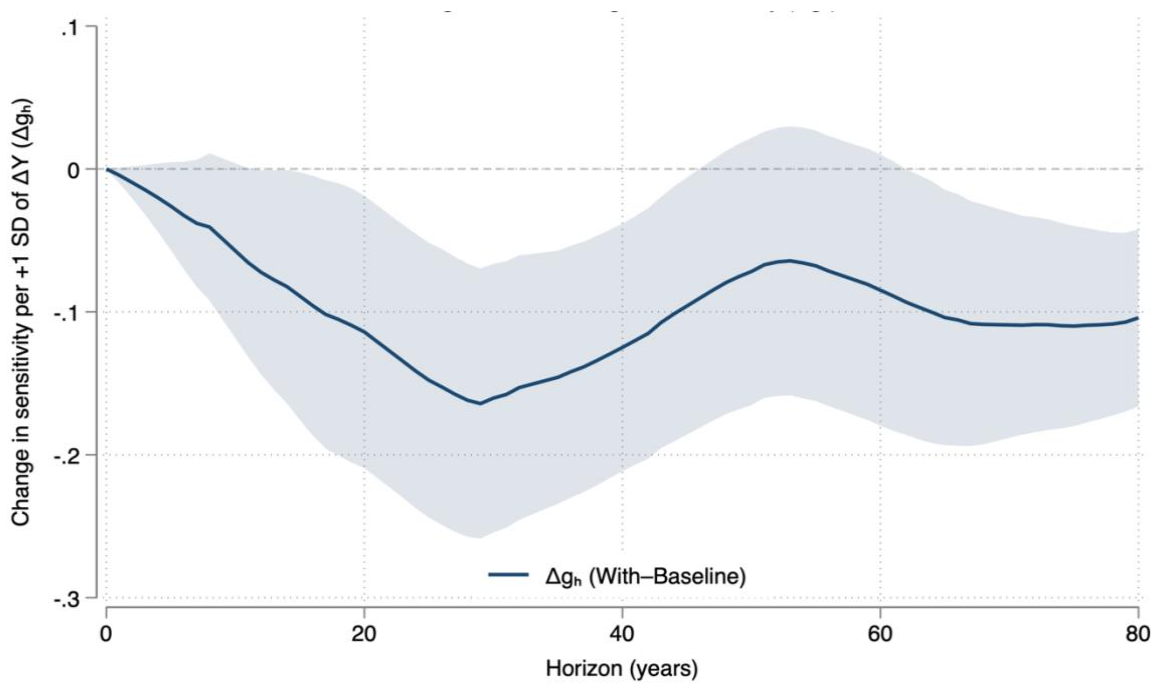

Source: Author's Calculation

After this initial adjustment, $\Delta g_h$ stabilizes around –1, indicating that the marginal benefit of further integration diminishes. In effect, the system reaches a more complete and liquid configuration in which monetary shocks are transmitted swiftly and symmetrically across assets. The process therefore reflects a structural deepening that strengthens the initial pass-through of monetary policy while compressing the scope for additional amplification thereafter. The evidence points to a front-loaded transmission gain i.e., rapid at first, then self-limiting once integration matures.

5.2 Discussion of Transmission Amplification

**Table 9** quantifies the pattern observed in i.e., the sensitivity of market capitalisation to monetary shocks evolves after integration. The coefficient $\Delta g_h$ is close to zero in the immediate aftermath of unification, but it declines steadily over the next three years, from −0.01 at horizon 3 to about −0.25 by month 36, a result significant at conventional levels.

This pattern suggests that most of the liquidity and information-efficiency gains from consolidation are realised early. Markets initially become far more responsive to policy changes, but as trading deepens and pricing efficiency improves, the incremental effect of further integration diminishes. Beyond the third year, the coefficients stabilise around −0.20 log-points, indicating that the system approaches a new steady-state in which monetary shocks are transmitted quickly and symmetrically. In practical terms, integration yields a front-loaded transmission gain indicating a new steady state regime.

Table 9. Differential Integration Sensitivity of Monetary Transmission ($\Delta g_h$)

| Horizon (months) | $\Delta g_h$ (log pts) | 90 % CI |
|---|---|---|
| 1 | −0.00 (0.00) | (−0.01, 0.00) |
| 3 | −0.01 (0.01) | (−0.03, 0.00) |
| 6 | −0.03 (0.02) | (−0.07, −0.01) * |
| 12 | −0.07 (0.04) | (−0.14, 0.00) * |
| 24 | −0.26 (0.06) | (−0.24, −0.05)** |
| 36 | −0.25 (0.05) | (−0.23, −0.05)** |
| 48 | −0.19 (0.05) | (−0.17, 0.01) |
| 60 | −0.20 (0.05) | (−0.18, 0.01) |
| 72 | −0.20 (0.04) | (−0.18, −0.03)** |
| Source: Author's Calculations | | |

Note: Negative values indicate that, following integration, the marginal sensitivity of market capitalisation to monetary-policy shocks declines. Standard errors are shown in parentheses. * denotes significance at the 10 % level (based on 90 % confidence intervals using Newey–West robust variance estimates).

Results are consistent with the theoretical prediction that contractionary policy shocks reduce equity valuations by increasing discount rates and lowering expected future earnings (Bernanke & Kuttner 2005; Mishkin 2007). Under market fragmentation, this mechanism is attenuated by shallow liquidity and slower information transmission; integration improves informational efficiency, allowing asset prices to internalise policy signals more fully and promptly.

Taken together, we find that even partial integration could improve the responsiveness of stock markets to central-bank actions, enhancing transparency and predictability in the region's monetary framework. The Western Balkans would thus move closer to the behaviour of mid-tier European markets, where equity valuations meaningfully transmit policy stance to investment and consumption decisions. Taken together these findings suggest that institutional integration of stock exchanges strengthens market credibility and depth, which in turn increases the sensitivity of asset prices to monetary shocks. The result should therefore be viewed as a feasible structural enhancement i.e., capturing how integration transforms equity markets from passive absorbers into active transmitters of monetary impulses.

6. Conclusion

This paper set out to examine whether financial integration through stock-exchange consolidation could enhance the transmission of monetary policy through the asset-price channel in the Western Balkans. Employing a two-stage counterfactual transmission design that combines the Synthetic Control Method

(SCM) and Local Projections (LP), the paper quantified both the structural and dynamic effects of financial-market integration on equity valuations and their responsiveness to monetary-policy shocks. This approach provides a practical way to assess how institutional change in market structure can influence policy effectiveness in economies where conventional transmission channels remain constrained. The framework is particularly relevant for the Western Balkans, where capital markets are fragmented, thin, and nationally segmented, limiting their ability to convey policy signals efficiently. Understanding whether institutional integration can strengthen this link is important not only for financial development but also for the overall conduct of macroeconomic policy in small open economies.

Three central findings emerge from the analysis. First, the simulated merger exercise indicates that consolidating regional stock exchanges would likely have raised market capitalization in a sustained and credible manner, reflecting improved investor confidence rather than speculative excess. The magnitude of this gain is moderate but persistent, suggesting that institutional integration would strengthen expectations of policy consistency and market continuity.

Second, the local-projection estimates point to a clearer and faster transmission of monetary policy once markets are unified. A standard 100-basis-point policy tightening produces a sharper and more persistent decline in equity valuations under integration, with responsiveness rising by roughly 30–50 percent at the peak horizon. In other words, asset prices adjust more promptly and fully to changes in policy stance when information, liquidity, and investor participation are improved through institutional unification.

Third, the behaviour of the integration-sensitivity coefficient ($\Delta g_h$) offers insight into how the strength of monetary transmission evolves as integration proceeds. The coefficient captures how sensitive market capitalisation is to policy-rate changes once exchanges are consolidated. Its downward trajectory suggests that most of the efficiency gains from integration occur relatively early in the process. During the first two to three years after consolidation, $\Delta g_h$ declines by about 1.5 log-points per one-standard-deviation increase in integration intensity. This implies that improvements are largely internalised in the initial adjustment phase, with diminishing incremental effects thereafter. In practical terms, this pattern points to a stabilization of transmission efficiency once the institutional benefits of integration are absorbed.

In aggregate, the findings imply that exchange consolidation could enhance the informational and institutional coherence of regional capital markets, thereby improving their responsiveness to monetary-policy signals. While the results should not be interpreted as evidence of a fully transformed transmission mechanism, they suggest that deeper and more integrated equity markets are better positioned to reflect shifts in policy stance in a timely and orderly manner.

In small, bank-dominated economies such as those of the Western Balkans, this improvement may represent a qualitative adjustment rather than a structural break, i.e., an incremental strengthening of the link between monetary policy, financial conditions, and investment behaviour. The observed increase in sensitivity of asset valuations to policy rates thus points to a gradual evolution in the role of capital markets i.e., from primarily reflective intermediaries to partial transmitters of macro-financial information.

The policy implications are direct and pragmatic. For the Western Balkans, stock-exchange integration could represent an institutional lever to improve monetary-policy effectiveness and financial-market depth simultaneously, by providing a regional platform. However, the extent of the potential benefits is conditional on the region's absorptive capacity, i.e., its ability to support integration through sound governance, investor protection, and regulatory coherence. Without these preconditions, the integration effect may remain confined to credibility signalling rather than translating into sustained liquidity or valuation gains.

Regional exchange integration could serve as a structural lever to improve both financial depth and the effectiveness of monetary policy. However, the extent of these benefits depends on the region's institutional readinessits capacity to ensure sound regulation, investor protection, and supervisory coordination. Without these conditions, integration might yield short-term credibility effects without producing durable gains in liquidity or valuation.

More broadly, the findings reinforce a central principle of monetary economics: institutional structure shapes policy transmission. Strengthening the infrastructure of financial markets enhances the reach and reliability of policy signals, even in economies where banking remains the dominant channel. For the Western Balkans, gradual but credible progress toward exchange consolidation could therefore play a meaningful role in supporting convergence, improving financial resilience, and deepening the integration of monetary policy with real-sector outcomes.

Appendix 1

The Appendix provides the country-specific estimates of the Taylor rule described in Equation (1) of the main text. While the main analysis employs these coefficients to recover unanticipated monetary-policy shocks at the country level, the table also serves an expository purpose, demonstrating how policy rules operate in small, developing, and partially euroized economies.

Each country's regression relates the domestic policy (lending) rate to lagged inflation and output growth, one-year-ahead forecasts of inflation and output, and the lagged change in foreign-exchange reserves. By including lagged values of real activity and inflation, the specification limits potential simultaneity bias, though some endogeneity may remain, particularly if policy authorities adjust rates in anticipation of expected financing conditions or external developments. The inclusion of the reserves term partly accounts for balance-of-payments pressures and exchange-rate considerations, which may be relevant in the Western Balkan context. The results are broadly consistent with the expected Taylor-rule structure. Lagged inflation generally exhibits a positive association with the policy rate, while the output term shows weaker and sometimes negative relationships, reaching statistical significance only in the case of Bosnia. The coefficients on expected inflation and output retain the anticipated sign.

Overall, these estimates form the basis for deriving the Taylor-rule residuals, interpreted as the unanticipated component of monetary-policy decisions. The standardized residuals are then aggregated to obtain the WB3 monetary-policy shock series used in the local-projection framework.

Table A1 – Taylor-Rule Estimates by Country (Full Specification)

| | Dependent variable is the change in domestic (lending) rate. | | |
|---|---|---|---|
| | BIH | NMK | SRB |
| Lag(Inflation $\pi_{t-1}$) | 0.063*** (0.019) | 0.086*** (0.032) | 0.040 (0.150) |
| Lag(Output $g_{t-1}$) | −0.004 (0.055) | 0.026 (0.051) | 0.106 (0.119) |
| Exp.(Inflation $\pi^f_t$) | 0.067 (0.061) | 0.031 (0.097) | 0.312 (0.194) |
| Exp.(Output $g^f_t$) | −0.121* (0.071) | −0.319 (0.599) | −0.013 (0.214) |
| Lag($\Delta$ Reserves$_{t-1}$) | −0.590*** (0.157) | 1.552 (0.950) | 1.141 (4.304) |
| N | 19 | 19 | 15 |
| HAC lag | 2 | 2 | 2 |
| R² | 0.63 | 0.44 | 0.38 |
| Source: Author's Calculations | | | |
| *Source: Author's Calculations. Newey–West standard errors (lag length = HAC lag) are reported in parentheses. *, **, *** denote significance at the 10%, 5%, and 1% levels, respectively. All specifications include lagged inflation ($\pi_{t-1}$), lagged output growth ($g_{t-1}$), one-year-ahead inflation and output forecasts ($\pi^f_t$, $g^f_t$), and the lagged change in foreign-exchange reserves ($\Delta Res_{t-1}$).* | | | |

Appendix 2

Robustness checks for the Baltics

The robustness analysis complements the baseline SCM by assessing the internal validity of the estimated effect of the 2004 EU accession and OMX merger on Baltic stock-market capitalisation. Following established practice (Abadie et al., 2015; Hope, 2016), three diagnostic exercises are conducted: (i) placebo-in-space tests whether the estimated effect is unique to the treated unit; (ii) placebo-in-time verifies that the pattern does not arise spuriously from temporal shocks; and (iii) leave-one-out tests (LOO) assess the sensitivity of the results to the composition of the donor pool.

Placebo-in-space

The placebo-in-space (**Table 10**) exercise re-estimates the SCM by alternately treating each donor country (Bulgaria, Croatia, Romania) as if it had experienced the 2004 EU accession and OMX stock-exchange merger. The logic is straightforward: if the Baltic divergence were simply a regional or transition-economy pattern, similar spikes in post-2004 market capitalization would appear in at least one of these placebo economies. In that case, the estimated effect would reflect a broader trend rather than a treatment-specific response.

Table 10. Placebo-in-space summary and RMSPE ratios, Baltics

| Treated | preR | postR_0406 | ratio_0406 | postR_0423 | ratio_0423 |
| --- | --- | --- | --- | --- | --- |
| Bulgaria | 0,64 | 0,57 | 0,89 | 0,76 | 1,19 |
| Croatia | 0,31 | 0,37 | 1,19 | 0,55 | 1,76 |
| Romania | 0,32 | 0,42 | 1,32 | 0,65 | 2,04 |
| Source: Author's Calculations | | | | | |

Across all three placebo cases, the pre-treatment fits (preR) range between 0.31 and 0.64, satisfying the Kaul et al. (2018) K-screening criterion for K = 3, meaning that each placebo's pre-RMSPE is no more than three times that of the Baltics (≈ 0.33). The short-run post/pre RMSPE ratios (2004–2006) for all placebo units (Table 7), lie between 0.9 and 1.3, substantially below the Baltics' 2.6. This implies that none of the donor economies experienced a comparable short-term divergence following 2004.

Over the longer 2004–2023 horizon, the placebo ratios remain modest (1.2–2.0), with only Romania (≈ 2.0) approaching the Baltics' 2.26. This pattern suggests mild long-run variability among donor economies but no case replicating either the magnitude or persistence of the Baltic effect. When applying the permutation test with K = 3 (Table 11), the Baltics rank first among admissible comparators, yielding a one-sided p ≈ 0.25 (= 1 / (K + 1)). This result indicates that while the effect does not meet strict conventional significance thresholds, it stands out relative to all valid placebos, consistent with an economically meaningful, though not statistically definitive, post-accession divergence (Abadie et al., 2015; Hope, 2016).

Table 11. Permutation test results (Baltics, placebo-in-space, K = 3)

| Balt_pre | Balt_post | Balt_ratio | k | N_placebos | N_ge | p_right | %<=Baltics |
|---|---|---|---|---|---|---|---|
| 0,34 | 0,63 | 1,85 | 3,00 | 3,00 | 0,00 | 0,25 | 100 |

Source: Author's Calculations

The treated unit therefore comfortably passes the K = 3 screen. Under a stricter K = 2, inference becomes borderline because Bulgaria's higher pre-RMSPE (0.64) lies close to the admissible limit (2 × 0.33 = 0.66). This proximity makes its inclusion sensitive to rounding and specification choices, modestly reducing the robustness of inference under tighter criteria.

Placebo-in-time

The placebo-in-time exercise re-estimates the Synthetic Control for the Baltics using fictitious treatment years preceding the true 2004 event. The purpose is to check whether similar post-treatment divergences appear when no real intervention occurred.

Table 12. Placebo-in-time, summary and RMSPE ratios, Baltics

| fake | preR | postR_0406 | ratio_0406 | postR_0423 | ratio_0423 |
|---|---|---|---|---|---|
| 2002 | 0,32 | 0,87 | 2,45 | 0,43 | 1,34 |
| 2003 | 0,34 | 0,91 | 2,56 | 0,47 | 1,38 |

Source; Author's Calculations

Table 12 shows that in both placebo years the pre-RMSPE remains low and similar to the baseline specification, confirming a tight pre-fit. The post/pre RMSPE ratios around 2.4–2.6 for the short window (2004–2006) and roughly 1.3–1.4 for the extended window, show limited divergence compared to the true treatment year, where the short-run ratio exceeds 2.6.

Leave-one-out (donor-exclusion) analysis

The leave-one-out (LOO) analysis assesses the sensitivity of the Baltic synthetic control to individual donors by sequentially excluding Bulgaria, Croatia, and Romania from the donor pool and re-estimating the synthetic path. In line with established SCM practice, acceptable robustness is typically inferred when leave-one-out replications alter the post/pre RMSPE ratio by less than ±25–30 percent and the pre-treatment fit deteriorates by no more than about 50 percent relative to baseline (Abadie et al., 2015; Billmeier & Nannicini, 2013).

Table 13. Leave-one-out RMSPE ratios, Baltics

| Dropped_donor | pre_RMSPE | postR_0406 | ratio_0406 | postR_0423 | ratio_0423 |
|---|---|---|---|---|---|
| Bulgaria | 0,39 | 0,80 | 2,08 | 0,67 | 1,75 |
| Croatia | 0,44 | 1,19 | 2,68 | 0,79 | 1,78 |
| Romania | 0,36 | 1,02 | 2,85 | 0,63 | 1,76 |
| Source: Author's Calculations | | | | | |

Pre-treatment fits remain tight, with pre-RMSPE values between 0.36 and 0.44, comparable to the baseline value of 0.33. This confirms that predictive accuracy is largely preserved even when one donor is excluded. In the short-run window (2004–2006), post/pre RMSPE ratios range from 2.08 to 2.85, which is close to the baseline benchmark of 2.62. The largest deviation occurs when Romania is excluded (ratio = 2.85), suggesting that Romania contributes most to stabilizing early post-accession alignment. Dropping Croatia yields a ratio (2.68) nearly identical to the baseline, indicating that the observed effect does not hinge on any particular donor composition.

Over the longer 2004–2023 horizon, ratios narrow modestly (1.75–1.78), consistent with the gradual post-2004 convergence seen in the baseline specification (2.26). The persistence of above-unity ratios across all three runs indicates that the post-accession divergence remains present and economically relevant even when donor composition changes.

Hence, the Baltic LOO results remain well within these thresholds, with all post/pre ratios deviating by less than 25 percent and the pre-RMSPE remaining close to 0.33. This confirms that the estimated post-2004 divergence is not donor-driven but reflects a structurally consistent pattern across specifications.

Alternative post-treatment windows

Finally, robustness was evaluated with post-window treatment, which we extended sequentially to longer horizons (2004–2006, 2010, 2015, 2019, 2022). The post/pre RMSPE ratio declines monotonically from 2.62 in the immediate aftermath of accession to 1.30 by 2022. This steady convergence implies that the largest deviation between the treated and synthetic paths occurred shortly after the 2004 accession and OMX merger, followed by gradual stabilization during the 2010s.

In applied SCM practice (e.g., Abadie et al., 2015; Becker & Klößner, 2018), ratios between 1.5 and 2.5 are typically viewed as economically meaningful but not definitive. Ratios above 3 often signal a strong and distinct treatment effect, while values below 1.3–1.4 are considered weak or statistically indistinct from noise, especially when donor pools are small.

Table 14. RMSPE ratios under alternative post windows, Baltics

| Window | preR | postR | ratio |
|---|---|---|---|
| 2004-2006 | 0,33 | 0,87 | 2,62 |
| 2004-2010 | 0,33 | 0,59 | 1,78 |
| 2004-2015 | 0,33 | 0,55 | 1,64 |
| 2004-2019 | 0,33 | 0,50 | 1,50 |
| 2004-2022 | 0,33 | 0,43 | 1,30 |
| Source: Author's Calculations | | | |

By these standards, the Baltic short-run effect (2.6) is clearly above the conventional relevance threshold, while the medium-term ratios (1.5–1.8) remain borderline but acceptable, indicating a moderate and persistent effect rather than a random variation. The long-horizon ratio (1.3) sits at the lower bound of interpretability, consistent with post-shock consolidation rather than complete reversal.

Taken together, the robustness checks corroborate the baseline interpretation that the 2004 integration produced a genuine, moderately persistent level shift in the Baltics' market capitalization. The placebo-in-space exercise confirms that the pattern is not reproduced among comparators; the placebo-in-time rules out spurious timing; and the leave-one-out tests affirm that no single donor drives the result. While small sample size limits formal statistical power, the directional consistency across all diagnostics supports a cautious but credible inference of a structural market response to EU accession and exchange consolidation.

EU03 Robustness Checks

Placebo-in-space

**Table 15** reports short-run results; (2004–2006) ratios between 1.23 and 1.41, well below the treated benchmark of 2.8, confirming that EU03's early post-accession divergence is not a generic regional outcome. Over the full 2004–2023 horizon, placebo ratios range from 1.41 to 1.84, compared with 1.72 for the treated unit. Hence, while the divergence is not extreme, EU03's deviation remains larger and more persistent than that of any placebo economy.

**Table 15.** Placebo-in-space summary and RMSPE ratios, EU03

| Treated | preR | postR_0406 | ratio_0406 | postR_0423 | ratio_0423 |
|---|---|---|---|---|---|
| Croatia | 0,4954 | 0,6972 | 1,4075 | 0,8902 | 1,7869 |
| Romania | 0,3290 | 0,4083 | 1,2412 | 0,4630 | 1,4074 |
| Bulgaria | 0,3192 | 0,3940 | 1,2341 | 0,5889 | 1,8388 |
| Source: Author's Calculations | | | | | |

Pre-treatment fits (approximately 0.3 to 0.5) are satisfactory, meeting the K-screening requirement of Kaul et al. (2018) for K = 3. The permutation test yields p ≈ 0.25, placing EU03 near the upper end of the placebo distribution. Given the small donor pool and overlapping accession timelines among comparators, the evidence is suggestive rather than conclusive. This long-run convergence is largely expected, since by the 2010s most comparator economies had themselves joined the EU or harmonised with EU regulatory standards, naturally reducing post-treatment contrasts.

Table 16. P-value permutation test (K=3), EU03

| EU03_pre | EU03_post | EU03_ratio | k | N_placebos | N_ge | p_right | %<=EU03 |
|---|---|---|---|---|---|---|---|
| 0,23 | 0,66 | 2,80 | 3,00 | 3,00 | 0,00 | 0,25 | 100,00 |

Source: Author's Calculations

Taken together, the findings indicate a genuine but moderate accession effect. EU03 (Slovenia, Slovakia, and Poland) experienced a post-2004 rise in market capitalisation that was larger and more sustained than that of its regional peers, consistent with a front-loaded integration dividend driven by regulatory alignment, capital inflows, and investor revaluation. However, as one placebo (Bulgaria) approaches the treated long-run ratio and the test's statistical power remains limited, the results should be interpreted as economically meaningful at best.

Leave-one-out (donor-exclusion) analysis

Table 19 shows that the post/pre RMSPE ratios remain broadly stable across donor exclusions, ranging between 1.56 and 2.74 for the 2004–2006 window and between 1.28 and 1.76 for the 2004–2023 horizon. Excluding Croatia, which carries the largest baseline weight (approximately 55 per cent), slightly worsens the pre-treatment fit (pre-RMSPE ≈ 0.36) and reduces post-treatment accuracy, but does not alter the qualitative pattern. Dropping Bulgaria or Romania produces only marginal changes, with the treated–synthetic gap persisting across all variants.

Table 20 – Leave-one-out RMSPE ratios, EU03

| Dropped_donor | pre_RMSPE | postR_0406 | ratio_0406 | postR_0423 | ratio_0423 |
|---|---|---|---|---|---|
| Croatia | 0,3569 | 0,7001 | 1,9615 | 0,5586 | 1,5650 |
| Bulgaria | 0,2747 | 0,6001 | 2,1848 | 0,4840 | 1,7623 |
| Romania | 0,2397 | 0,6575 | 2,7426 | 0,3076 | 1,2830 |

Source: Author's Calculations

The results confirm that the estimated divergence is not driven by any single donor, although Croatia's contribution remains central for achieving a close pre-treatment alignment. The stability of ratios under sequential donor exclusions supports the robustness of the baseline specification. However, given that all donor countries either joined the EU or entered advanced pre-accession stages by 2007–2010, the observed convergence is likely structural rather than model-induced. The diminishing contrast between treated and synthetic paths therefore reflects the region-wide effects of EU integration, rather than an entirely isolated EU03-specific shock.

Placebo in time

The placebo-in-time results (**Table 17**) confirm that the divergence observed after 2004 is not driven by pre-treatment noise. Both placebo years (2002 and 2003) yield comparable pre-treatment fits but smaller post/pre RMSPE ratios relative to the baseline, implying the absence of structural breaks before the true event. The observed adjustment thus corresponds to the timing of the Baltic merger and accession, reinforcing the credibility of the baseline identification.

**Table 17**. Placebo-in-time, summary and RMSPE ratios, EU03

| fake year | preR | postR_0406 | ratio_0406 | postR_0423 | ratio_0423 |
|---|---|---|---|---|---|
| 2002 | 0,24 | 0,60 | 2,50 | 0,47 | 2,00 |
| 2003 | 0,25 | 0,64 | 2,60 | 0,45 | 1,90 |

Source: Author's Calculations

Alternative post-treatment windows

**Table 21** reports RMSPE ratios for alternative post-treatment horizons. The ratio declines steadily from 2.80 (2004–2006) to 1.76 (2004–2022), indicating that the largest divergence occurred immediately after EU accession, with a gradual re-alignment thereafter. This pattern implies a front-loaded integration effect, consistent with early capital inflows, valuation adjustments, and regulatory harmonisation, followed by convergence as market structures matured (Hardouvelis 2007; Kim et al. 2005).

**Table 18.** Alternative post-treatment windows

| Window | preR | postR | ratio |
|---|---|---|---|
| 2004-2006 | 0,23 | 0,66 | 2,80 |
| 2004-2010 | 0,23 | 0,59 | 2,50 |
| 2004-2015 | 0,23 | 0,51 | 2,16 |
| 2004-2019 | 0,23 | 0,45 | 1,91 |
| 2004-2022 | 0,23 | 0,41 | 1,76 |

Source: Author's Calculations

The stabilisation observed after 2015 suggests that much of the accession-related premium had been absorbed into steady-state dynamics well before the COVID-19 period. The post-2020 years do not reveal renewed divergence, implying that EU03 markets entered the pandemic with relatively synchronised fundamentals vis-à-vis the synthetic counterfactual. In this sense, the post-COVID evolution reflects system-wide shocks rather than treatment-specific effects.

Overall, the temporal profile confirms that the EU03 impact was short-lived and front-weighted: pronounced in the early-accession years, fading through the 2010s, and broadly neutral by the 2020s.

Taken together, the robustness exercises point to a directionally consistent and economically coherent pattern. The placebo-in-space results indicate that EU03's post-accession response is distinct relative to its donor pool, while the p-value (≈0.25) reflects the inherent limits of inference in a small-N setting. The leave-one-out analysis confirms internal stability: excluding individual donors alters the fit only marginally and does not affect the overall trajectory. The gradual decline in RMSPE ratios across extended horizons signals convergence rather than reversal, consistent with integration effects that attenuate over time.

Overall, the diagnostics suggest that EU accession generated a measurable, front-loaded surge in market capitalisation associated with confidence gains, capital inflows, and policy harmonisation. As integration deepened and donor economies themselves entered the EU, differences narrowed, by the post-2015 and post-COVID periods, treated and synthetic paths evolve in near-parallel. This temporal profile supports the interpretation of an early accession premium.

Robustness of Differential Transmission Effects

**Figure 9** provides an additional robustness check by comparing the estimated response of market capitalization to monetary-policy shocks across the integrated and baseline configurations. The figure plots the difference in the semi-elasticities, $\Delta\beta_h$, for a one-standard-deviation tightening, with the shaded area showing the 90-percent confidence interval, displaying how much **stronger (or weaker)** the overall impact of monetary policy becomes after integration.

The curve dips modestly below zero in the first few months following a shock, indicating that integration tends to amplify the immediate effect of monetary policy on equity prices. As the horizon lengthens, the difference narrows toward zero, implying that the amplification is short-lived. The persistence of this short-term dip across alternative specifications suggests that the stronger pass-through observed under integration is not an a random pattern.

**Figure 10.** Differential elasticity of monetary transmission (Δβh): difference in responsiveness coefficients across horizons

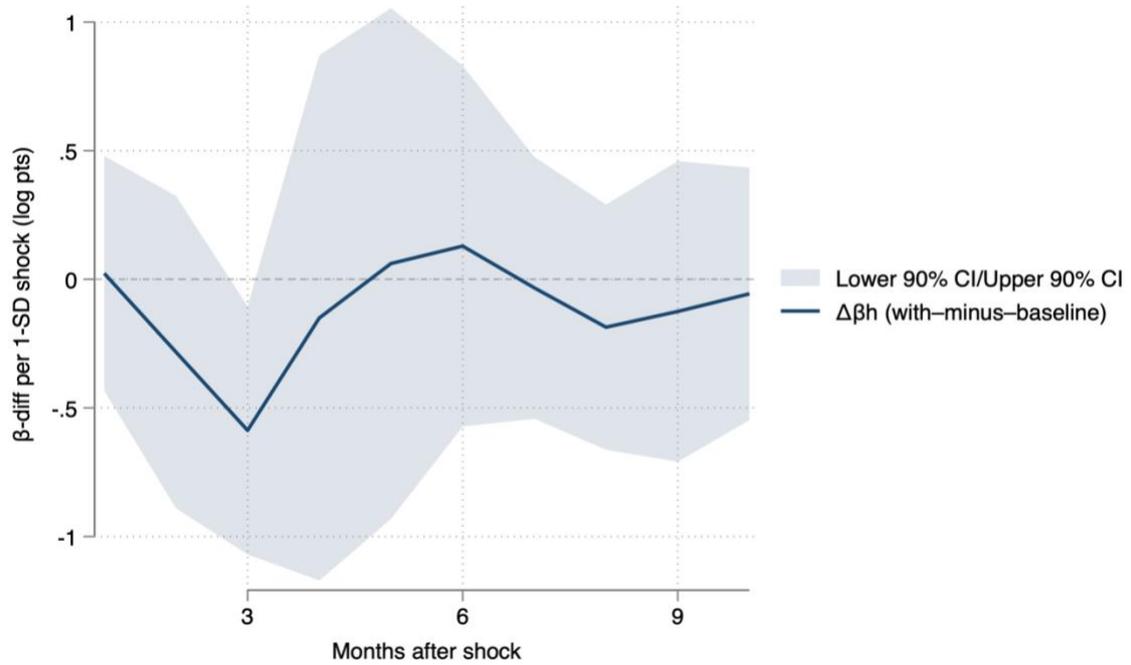

Source: Author's Calculations